\documentclass[useAMS,usenatbib]{mn2e}

\usepackage{color}
\pdfoutput=1	%for ARXIV
\usepackage{graphicx}
\title[Trumpler 5]{The old, metal-poor, anticentre open cluster Trumpler 5\thanks{This work is based on observations made at the ESO telescopes under programmes 68.D-0212 and 074.D-0344}}
\author[Donati et al.]
{P. Donati$^{1,2}$,
G. Cocozza$^{1}$, 
A. Bragaglia$^{1}$,
E. Pancino$^{1,3}$,
T. Cantat Gaudin$^{4,5}$,
\newauthor
R. Carrera$^{6.7}$, 
M. Tosi$^{1}$\\
$^{1}$INAF-Osservatorio Astronomico di Bologna, via Ranzani 1, 40127 Bologna, Italy\\
$^{2}$Dipartimento di Fisica e Astronomia, Universit\`a di Bologna, via Ranzani 1, 40127 Bologna, Italy\\
$^{3}$ASI Science Data Center, I-00044 Frascati, Italy\\
$^{4}$Dipartimento di Fisica e Astronomia, Universit\`a di Padova, vicolo Osservatorio 3, 35122 Padova, Italy\\
$^{5}$INAF-Osservatorio Astronomico di Padova, vicolo Osservatorio 5, 35122 Padova, Italy \\
$^{6}$Instituto de Astrof\'{\i}sica de Canarias, La Laguna, Tenerife, Spain \\
$^{7}$Departamento de Astrof\'{\i}sica, Universidad de La Laguna, Tenerife, Spain
}

\begin{document}

\date{}

\pagerange{\pageref{firstpage}--\pageref{lastpage}} \pubyear{2013}

\maketitle

\label{firstpage}

\begin{abstract}
As part of a long term programme, we analyse the evolutionary status and
properties of the old  and populous open cluster Trumpler~5 (Tr~5), located in the
Galactic anticentre direction, almost on the Galactic plane. Tr~5 was observed
with WFI@MPG/ESO Telescope using the Bessel $U$, $B$, and $V$ filters. The
cluster parameters have been obtained using the synthetic colour-magnitude
diagram (CMD) method, i.e. the direct comparison of the observational CMD with a
library of synthetic CMDs generated with different stellar evolution sets (Padova,
FRANEC, and FST). Age, reddening, and distance are derived through the synthetic CMD method using stellar evolutionary models with subsolar metallicity ($Z=0.004$ or $Z=0.006$). Additional spectroscopic observations with UVES@VLT of three
red clump stars of the cluster were used to determine more robustly the chemical properties of the
cluster. Our analysis shows that Tr~5 has subsolar metallicity, with [Fe/H]$=-0.403\pm0.006$ dex (derived from spectroscopy), age between 2.9 and 4 Gyr (the lower age is found using
stellar models without core overshooting), reddening $E(B-V)$ in the range 0.60
to 0.66 mag complicated by a differential pattern (of the order of $\sim\pm0.1$
mag), and distance modulus $(m-M)_0=12.4\pm0.1$ mag. 
\end{abstract}


\begin{table*}
\centering
\caption{Logbook of the WFI observations (RA and Dec refer to the centre of the telescope pointings).}
\label{tab:logbook}
  \begin{tabular}{@{}lcclcccl@{}}
  \hline
   Name   &   RA (h m s)  & Dec ($\degr$ $\arcmin$  $\arcsec$) & UT date & U & B & V & seeing \\
        & J2000 & J2000 &  & Exp. time~(s) & Exp. time~(s) &   Exp. time~(s)  & $\arcsec$   \\
 \hline
 
Tr~5& 6 36 16 & 9 22 50 & 25 Nov 2001 & 10, 600 & 3, 600 & 3, 600 & 0.9-1.3 \\
Tr~5 ext& 6 36 16 & 8 32 50 & 25 Nov 2001 & 10, 600 & 3, 600 & 3, 600 & 0.9-1.3 \\
\hline
\end{tabular}
\end{table*}

\section[]{Introduction}\label{intro}
Old open clusters (OCs) are ideal probes of the Galactic disc structure, formation, and chemical distribution \citep[see e.g.,][]{friel95,bt06,
magrini09,pancino10,lepine11}.   
Old clusters are only $\sim15$\% of the whole population of more than 2100 known OCs \citep[see][]{dias02} and only $\sim5$\% are older than 2~Gyr. However, old OCs are particularly important to  constrain the formation and evolution, both dynamical and chemical, of the Milky Way disc.
Using a combination of photometry, spectroscopy, and models, OCs' fundamental parameters (age, distance, and metallicity) are relatively easy to determine. In the Gaia era OCs will retain  their importance. In fact, Gaia will produce exquisitely precise distances and proper motions for (almost) all MW OCs. This will permit, for instance,  to isolate the true cluster members,  producing the best templates of simple stellar populations of different ages (hence stellar masses) to be used as robust test of all details of stellar evolutionary models. In turn, this will produce the best age estimates and will permit to derive the fine details of the evolution of the disc.

In the BOCCE (Bologna Open Cluster Chemical Evolution) project \citep[][and references therein]{bt06,donati14a} we use both the comparison between observed colour-magnitude diagrams (CMD) and stellar evolutionary models and the analysis of high-resolution spectra of stars to infer the cluster properties. We present here our study of
Trumpler~5 (Tr~5 hereafter), a massive, old OC in the anticentre direction ($l=202.865\degr$, $b=1.050\degr$, \citealt{dias02} and web updates).  Tr~5 is interesting because it
is metal-poor (${\rm [Fe/H]\le -0.3}$ dex, according both to photometric and spectroscopic measures) and located  at a Galactocentric distance ($R_{GC}$) of 10-13 kpc where a transition from a radially-decreasing metallicity to a flat(ter) distribution seems to be present \cite[e.g.,][and references therein]{yong12}.
 
As is often the case for OCs, the properties of Tr~5  (age, distance, reddening) as measured in the literature show a large dispersion \citep[see Table 1 in][for a comprehensive list of values prior to the present paper]{kks09}, even if the most recent determinations seem to agree better. The first photographic photometry of Tr~5 by \cite{dh70} did not reach the main sequence (MS) 
turn-off (TO), but already showed that the cluster is rich, old, and located in a highly (and differentially) reddened region. \citet[][K98 hereafter]{kaluzny98} obtained CCD $BVI$ photometry with the 0.9m and 2.1m telescopes on Kitt Peak; his data show a very well defined MS, a red giant branch (RGB), and an elongated red clump (RC). By comparing Tr~5 to M67, K98 found it only slightly younger. His distance modulus was $(m-M)_0=12.4$ mag and the reddening is $E(B-V)=0.58$ mag, assuming solar metallicity. K98 found Tr~5 to be very massive (at least 3000 M$_\odot$). Using Kaluzny's data but fitting them with isochrones, \cite{ks03} found a younger age (2.4 Gyr), a consistent distance modulus and reddening (12.64 mag and 0.64 mag, respectively), and [Fe/H]=-0.30 dex.
\citet[][P04, hereafter]{piatti04}, on the basis of $VI_C$ and Washington photometry obtained with the 0.9m telescope at Cerro Tololo and comparison with theoretical isochrones, determined rather different results for age (5 Gyr) and distance ($(m-M)_V=13.8$ mag), while agreeing with past studies on reddening ($E(V-I)=0.80$ mag, i.e., $E(B-V)=0.64$ mag) and metallicity.
They also determined a differential reddening ($\Delta E(B-V)=0.11-0.22$ mag) and measured a cluster radius of about 7.7 arcmin. Finally, \cite{kks09} used the infrared 2MASS data and stellar isochrones to determine an age of 2.8 Gyr, an average reddening $E(B-V)=0.64$ mag,  a distance modulus $(m-M)_0=12.64$ mag, and [Fe/H]=-0.4 dex.

The metallicity of Tr~5 seems to be the least  controversial parameter, since all studies agree on a subsolar value. This is confirmed by spectroscopy. Two independent studies employed the infrared calcium triplet (CaT). \cite{cole04} used a set of globular clusters (GCs) and OCs to calibrate the CaT method on literature data and applied the derived calibration to Tr~5, for which they obtained spectra of 16 stars (14 of which are radial velocity members, with an average
value of $54\pm5$ km~s$^{-1}$). They obtained
[Fe/H]$=-0.56\pm0.11$ dex, making Tr~5 one of the metal-poorer known OC. This value was revised upwards by \cite{carrera07}, who found instead [Fe/H]$=-0.36\pm0.05$ dex, from the spectra of 17 member stars (average velocity $44\pm10$ km~s$^{-1}$) and their CaT metallicity calibration based on many GCs and OCs. Finally, in a paper dedicated to the analysis of one lithium-rich evolved star, \cite{monaco} found [Fe/H]=$-0.49$ (rms 0.04) dex from four giants observed with the UVES-FLAMES and MIKE spectrographs.

We present here our results obtained with the synthetic CMD
technique for age, distance, and reddening and with the analysis of high-resolution
spectra for metallicity and elemental abundances. The paper is organised as
follows. The photometric and spectroscopic observations are described in Sec.~\ref{sec:data}, the spectroscopic analysis is illustrated in Sec.~\ref{abu}. Sec.~\ref{CMDsynth} is dedicated to the photometric analysis of the cluster evolutionary properties. Summary and conclusions are in Sec.~\ref{summary}.

\begin{table}
 \centering
\caption{Completeness of our photometry expressed in percentage. 
For magnitudes brighter than 16 completeness is 100\%.}
\label{tab:compl}
  \begin{tabular}{@{}lcc@{}}
  \hline
mag & compl $B$ & compl $V$ \\%  & compl $U$
  \hline
16.0 & 100 $\pm$  0.1  & 100 $\pm$  0.1 \\%
16.5 & 100 $\pm$  0.1  &  99 $\pm$  0.2 \\%
17.0 & 100 $\pm$  0.1  &  98 $\pm$  0.2 \\%
17.5 & 100 $\pm$  0.2  &  98 $\pm$  0.3 \\%
18.0 &  98 $\pm$  0.3  &  98 $\pm$  0.4 \\%
18.5 &  98 $\pm$  0.3  &  97 $\pm$  0.5 \\%
19.0 &  98 $\pm$  0.4  &  97 $\pm$  0.7 \\%
19.5 &  97 $\pm$  0.6  &  95 $\pm$  1.0 \\%
20.0 &  97 $\pm$  0.8  &  95 $\pm$  1.4 \\%
20.5 &  96 $\pm$  1.1  &  93 $\pm$  2.1 \\%
21.0 &  95 $\pm$  2.5  &  83 $\pm$  3.2 \\%
21.5 &  95 $\pm$  2.2  &  65 $\pm$  5.0 \\%
22.0 &  93 $\pm$  3.2  &  41 $\pm$  7.0 \\%
22.5 &  86 $\pm$  4.6  &  12 $\pm$ 10.4 \\%
23.0 &  74 $\pm$  6.9  &   2 $\pm$ 13.3 \\%
23.5 &  63 $\pm$ 10.6  &   \\
24.0 &  50 $\pm$ 16.0  &   \\
24.5 &  28 $\pm$ 23.2  &   \\
25.0 &  07 $\pm$ 30.1  &   \\
25.5 &  01 $\pm$ 33.6  &   \\
\hline
\end{tabular}
\end{table}

\begin{figure}%
\centering
\includegraphics[scale=0.9]{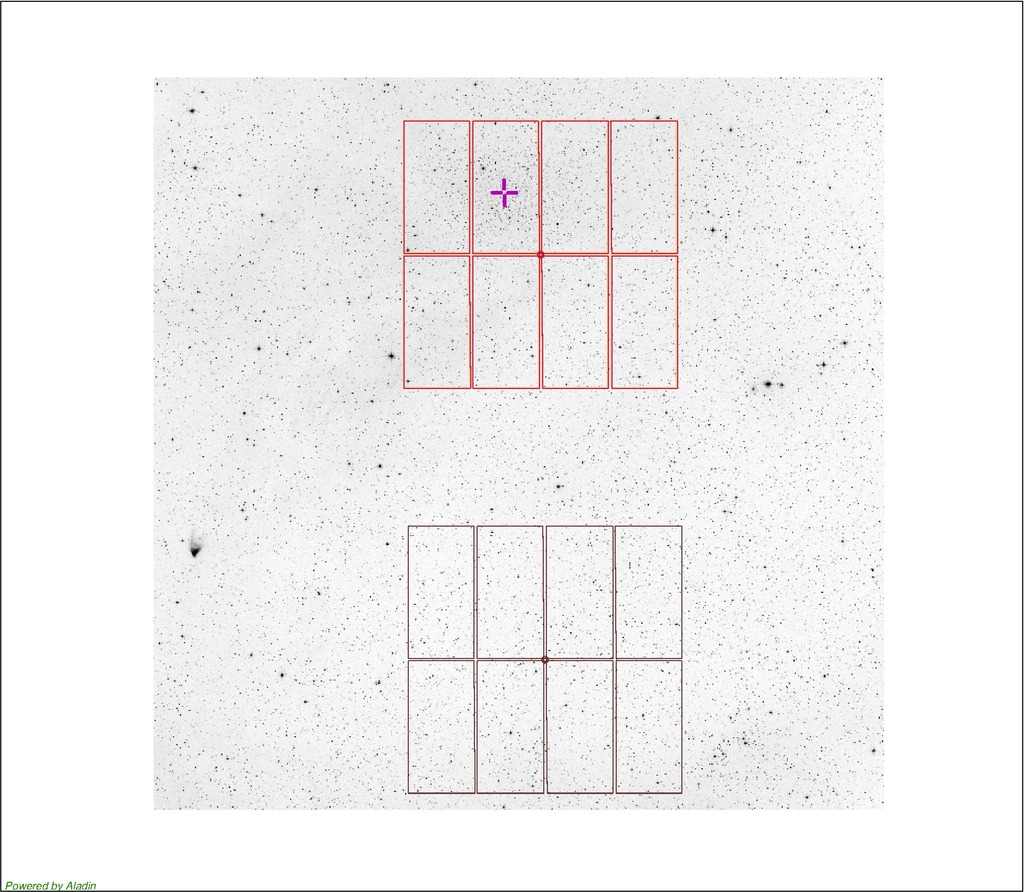} 
\caption{DSS image with the footprint of the WFI field of view for Tr~5 (upper) 
and the comparison field (lower). CCD~\#1 is the upper, left-most; CCD numeration is clockwise. The cluster centre is indicated by a cross on CCD~\#2. North is up, East is left. }
\label{fig:pointing}
\end{figure}

\section[]{Observations and Data Analysis}\label{sec:data}
\subsection{Imaging observations}
Observations of Tr~5  were performed with the Wide Field Imager mounted at the 
2.2-m MPG/ESO Telescope of the La Silla Observatory (Chile), in November 2001. 
The WFI instrument is a mosaic composed by two rows of four EEV CCDs
(2096$\times$4098 pixels) with a total field of view (FoV) of the instrument of
34$\times$33 arcmin$^{2}$,  with a pixel scale of 0.238
arcsec/pixel. 

The dataset consists of one long and two short exposures  in each of the U, B,
and V filters. The logbook of the observation is given in
Table~\ref{tab:logbook}. The seeing was better than 1.3$\arcsec$ for all the
images. In
Fig.~\ref{fig:pointing}  \citep[produced with  Aladin,][]{aladin} we show a  Digitized Sky Survey (DSS) image of our field,
with the eight WFI CCDs indicated.  The cluster is almost entirely located in 
Chip \#2 of the mosaic. One external field was also observed, in order to take
into account the background/foreground contamination, pointing the telescope
50$\arcmin$ south of the cluster centre. The standard field SA92
\citep{landolt} was also observed for photometric calibration, but the night was  
probably not photometric due to thin veils scattered
all over the sky (see below for the calibration to the standard system).

\subsection{Imaging data reduction}
The raw WFI images were corrected for bias and flat field, 
by using the standard package \emph{mscred} included in  IRAF.\footnote{IRAF is the Image Reduction
and Analysis Facility, distributed by the National Optical Astronomy Observatory,which is operated by the 
Association of Universities  for Research in Astronomy (AURA) under cooperative agreement with the 
National Science Foundation.}
The source detection and the instrumental magnitudes were performed
independently for each U, B, and V frame and for each CCD of the WFI mosaic,
using the Point Spread Function (PSF) fitting code {\sc daophot ii/allstar}
\citep{ste1,ste2}.  For each CCD a selected sample (120-170) of bright and
well isolated stars was adopted to compute the PSF in each exposure. 
In order to minimise geometrical distortions, which are present in the
WFI mosaic, we used a spatially variable PSF, with a
quadratic dependence on both \emph{x} and \emph{y} coordinates.  
We used the \emph{2 Micron All Sky Survey}  Catalogue (2MASS, \citealt{2mass}) to compute the astrometric
solution and transform the instrumental pixel coordinates into J2000
celestial coordinates. More than 2000 2MASS stars were used as astrometric
standards, and  cross-correlated with our catalogue using the CataXcorr
code\footnote{http://davide2.bo.astro.it/$\sim$paolo/Main/CataPack.html},  developed by P. Montegriffo and widely used  by our group in the past 10 years. The r.m.s. scatter
of the solution was $\sim 0.1 \arcsec$ in both RA and Dec. 

In order to derive
the completeness level of our photometry we used the procedure 
successfully adopted in several previous studies by our group 
\citep[e.g.][for a description]{donati12}.
About $\sim$250000 stars have been artificially  created and added uniformly
in the deepest frames, in groups of $\sim$ 900 stars at each step, thus
mimicking the addition of one single star at each time and avoiding any impact on the  actual crowding condition. The instrumental
magnitudes of all the artificial stars have been recovered using the same
strategy described above, computing the completeness level of our photometry
as the fraction of stars recovered for each magnitude interval. The result is
shown in Table~\ref{tab:compl} but  only for the $B$ and $V$ magnitudes, the two passbands we were able to calibrate (see next section).

\subsection{Photometric calibration}
We intended to tie our instrumental magnitudes to the Johnson-Cousins photometric system using the observations of photometric standard stars. 
However, we encountered problems, due both to
the non photometric conditions during the observations and to  
the paucity of standard stars in each single CCD, which forced us to use one single calibration equation for the whole mosaic in each filter. In fact, a
comparison with photometric catalogues available in literature (K98, P04) clearly exhibited systematic differences of our preliminary calibration varying  from CCD to CCD (differences up to 0.5 mag in the $B-V$ colour). 
We supposed that this could be due to the well-known illumination problem of WFI, which was investigated by \citet{koch03}. We therefore followed their same approach to correct the instrumental magnitudes for this effect but found that the overall correction was  within $\pm0.1$ mag in colour, too small with respect to the systematic differences we had from CCD to CCD. We concluded that in our case the illumination was only part of the problem and that we really needed a calibration equation for each CCD. 

\begin{figure}
\centering
\includegraphics[scale=0.9]{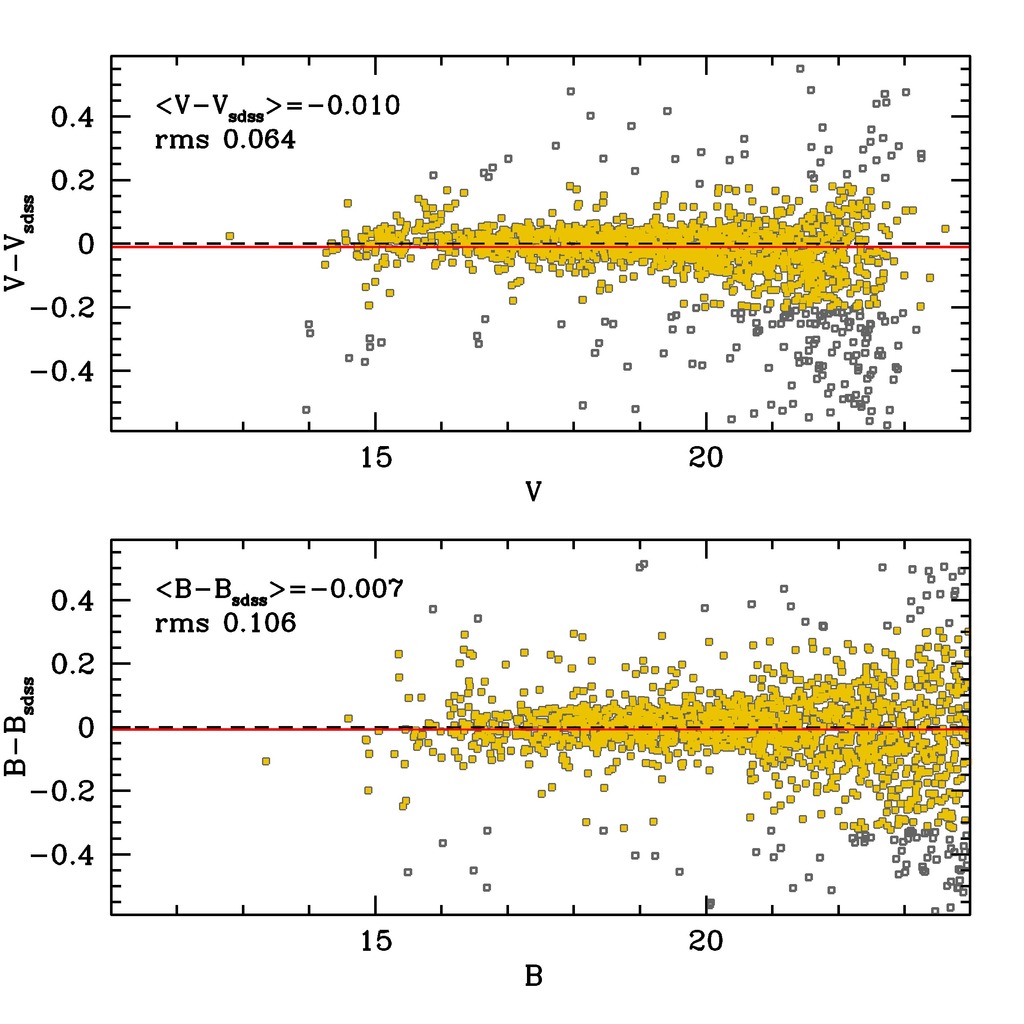} 
\caption{Comparison of the calibrated $B$ and $V$ magnitudes with the SDSS photometry for CCD~\#2. The yellow-filled dots
are the stars used to compute the mean difference within 2$\sigma$ from the
average. The differences are compatible with 0, without trends with magnitude. The same conclusions apply to all the other CCDs of the instrument.}
\label{fig:equcomp}
\end{figure}

Therefore, we used the photometric catalogue of the stars observed by the Sloan Digital Sky Survey (SDSS) as a sort of secondary standards to obtain an independent calibration equation for each of the eight CCDs and we transformed the SDSS magnitudes to the Johnson-Cousins system\footnote{For the conversion we used the equations available at http://www.sdss.org/dr4/algorithms/sdssUBVRITransform.html\#Lupton2005}. 
The same procedure was also applied to the external field. 
Unfortunately, it was possible to calibrate only the $B$ and $V$ images and not the $U$ data because there are no transformations available for this passband. The calibration equations used for the FoV centred on the cluster are summarised in Table~\ref{tab:equ}; the comparisons with the original SDSS catalogue are shown in Fig.~\ref{fig:equcomp} for CCD~\#2.
For this CCD, in particular, we corrected the calibration of the $B$ magnitude with a second iteration because a colour term was still present in the comparison with the SDSS catalogue (this occurrence did not show up for the external field). 

\begin{table}
 \centering
\caption{Calibration equations obtained for the eight CCDs for the Tr~5 pointing.}
\begin{tabular}{ccc}
\hline
\hline
\multicolumn{3}{c}{CCD 1} \\
\hline
equation & rms & stars used\\
\hline
$B-b=24.576+0.253\times(b-v)$ & rms 0.01 & about 1000\\
$V-v=24.010-0.053\times(b-v)$ & rms 0.04 & about 1200\\
\hline
\multicolumn{3}{c}{CCD 2$^{a}$} \\
\hline
equation & rms & stars used\\
\hline
$B-b=24.533+0.355\times(b-v)$ & rms 0.04 & about 1000\\
$V-v=24.026-0.110\times(b-v)$ & rms 0.03 & about 1200\\
\hline
\multicolumn{3}{c}{CCD 3} \\
\hline
equation & rms & stars used\\
\hline
$B-b=24.511+0.312\times(b-v)$ & rms 0.04 & about 1600\\
$V-v=24.016-0.080\times(b-v)$ & rms 0.04 & about 2000\\
\hline
\multicolumn{3}{c}{CCD 4} \\
\hline
equation & rms & stars used\\
\hline
$B-b=24.534+0.316\times(b-v)$ & rms 0.05 & about 1000\\
$V-v=24.010-0.001\times(b-v)$ & rms 0.03 & about 1000\\
\hline
\multicolumn{3}{c}{CCD 5} \\
\hline
equation & rms & stars used\\
\hline
$B-b=24.564+0.278\times(b-v)$ & rms 0.04 & about 1000\\
$V-v=24.021-0.115\times(b-v)$ & rms 0.04 & about 1200\\
\hline
\multicolumn{3}{c}{CCD 6} \\
\hline
equation & rms & stars used\\
\hline
$B-b=24.504+0.387\times(b-v)$ & rms 0.04 & about 1000\\
$V-v=23.961-0.042\times(b-v)$ & rms 0.03 & about 1600\\
\hline
\multicolumn{3}{c}{CCD 7} \\
\hline
equation & rms & stars used\\
\hline
$B-b=24.496+0.379\times(b-v)$ & rms 0.05 & about 1000\\
$V-v=23.953-0.053\times(b-v)$ & rms 0.04 & about 1400\\
\hline
\multicolumn{3}{c}{CCD 8} \\
\hline
equation & rms & stars used\\
\hline
$B-b=24.545+0.272\times(b-v)$ & rms 0.04 & about  800\\
$V-v=23.975-0.070\times(b-v)$ & rms 0.03 & about 1000\\
\hline
\end{tabular}
\label{tab:equ}

$^aB$ magnitude obtained from this equation is then corrected to compensate for a colour term using the following equation: $B^*=-0.012\times B+0.237$.  
\end{table}

The final catalogue contains $B$ and $V$ Johnson-Cousins magnitudes for 39660 objects and is made available through the Centre de Donn\'{e}es de Strasbourg (CDS) and the web database WEBDA\footnote{http://webda.physics.muni.cz}.

\subsection{Comparison with previous data}
Tr~5 was previously studied by many authors, as mentioned in Sect.~\ref{intro}. In particular the photometric 
catalogues derived by K98 and by P04 are available in the WEBDA. 

\begin{figure}
\includegraphics[scale=0.9]{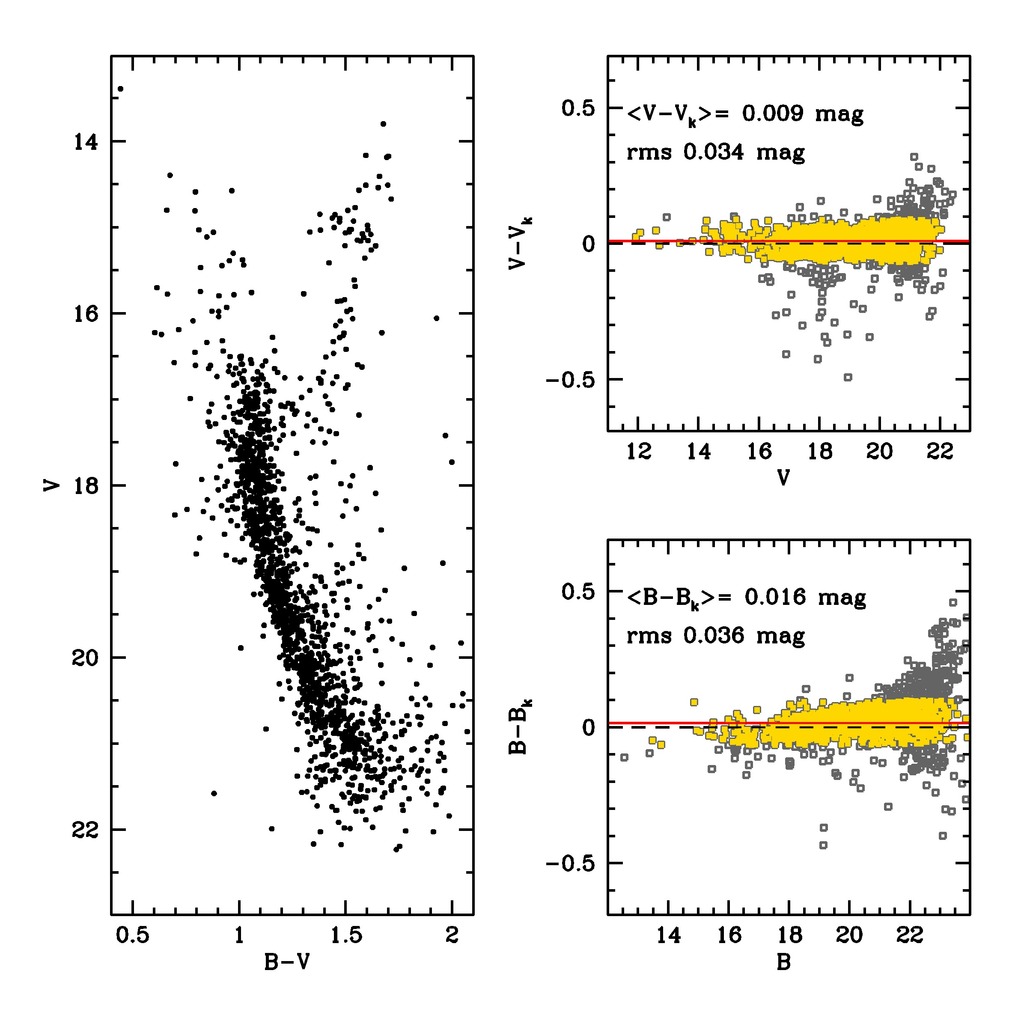}
\caption{Left panel: CMD of Tr~5 in the $V,BV$ plane from run 2 by
K98. Right panels: Differences between our photometry and the K98 one in $V$ (upper panel) and $B$ (bottom panel). The points in the right panels show the data for all the stars in common, while the yellow-filled dots
are the stars used to compute the mean difference within 2$\sigma$ from the
average.}
\label{Diff_CK}
\end{figure}

Of the three  datasets in K98  
(the first using the $V,I$ filters, and the others using the $B,V$ filters, see his Table~1), we decided to use for 
the comparison with our photometry only the second run,  obtained in better conditions and with a better instrument.
Moreover the FoV of this dataset nicely matches the FoV of CCD~\#2 of the WFI instrument,
which includes almost completely Tr~5.
In Fig.~\ref{Diff_CK} we show the differences between our photometry and the photometry by K98. In general $B$ and $V$ compare well; differences are within 0.02 mag with a mild dependence on magnitude, especially in $B$. Our bright stars are in general slightly bluer than in K98. This is probably due to the fact that the $B$ filter mounted on the WFI instrument differs from the classical
$B$ filter in the Johnson-Cousins system used by K98. 

\begin{figure}
\includegraphics[scale=0.9]{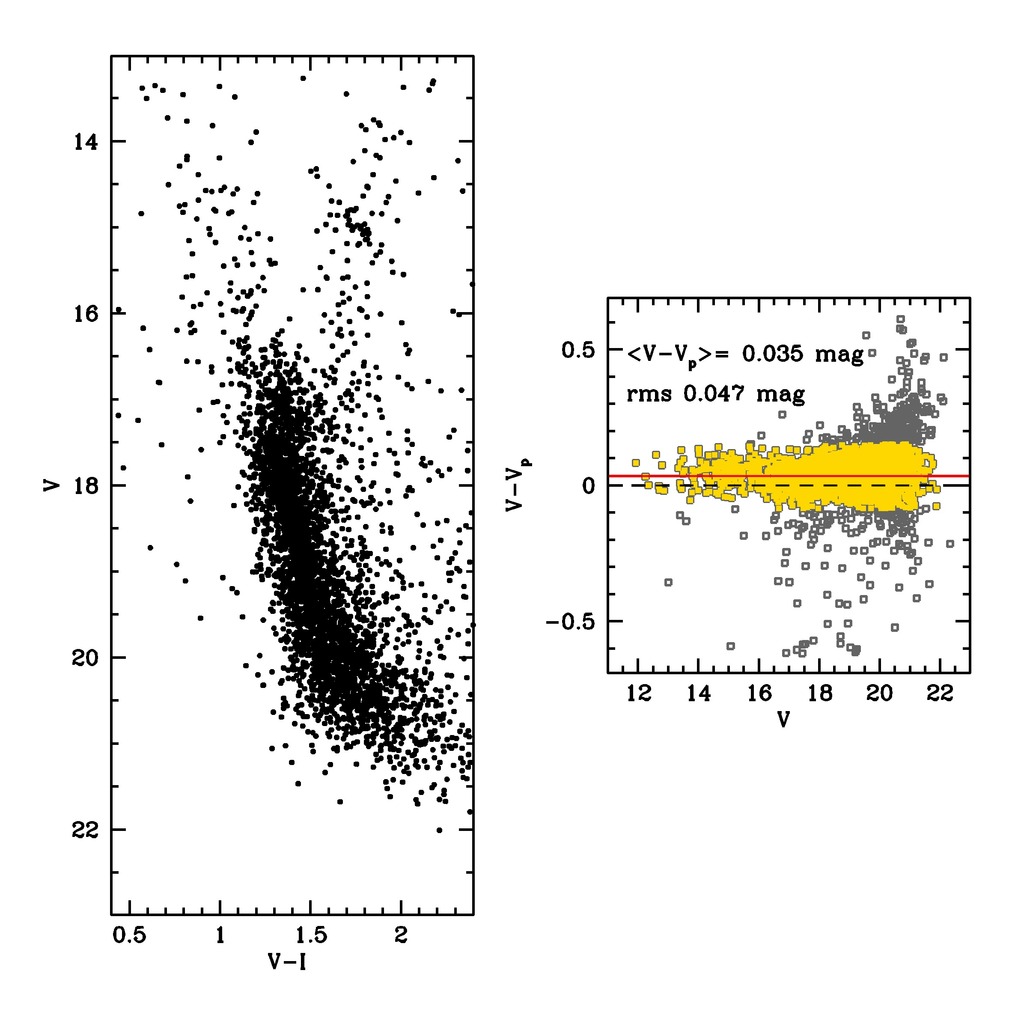}
\caption{Left panel: CMD of Tr~5 in the $V,VI$ plane from
P04. Right panel: Differences between our photometry and the
P04 one in $V$.}
\label{Diff_CP}
\end{figure}

In the case of P04 (see Fig.~\ref{Diff_CP}), only the $V$ magnitude can be compared. We found an average difference of $<V-V_P>=0.035$ mag, larger than the difference found with the $V$ of K98 but with no evident trend with magnitude. 

We deemed both comparisons acceptable and took them as an indication of the validity of our calibration on the SDSS data.

\begin{figure}
\includegraphics[scale=0.9]{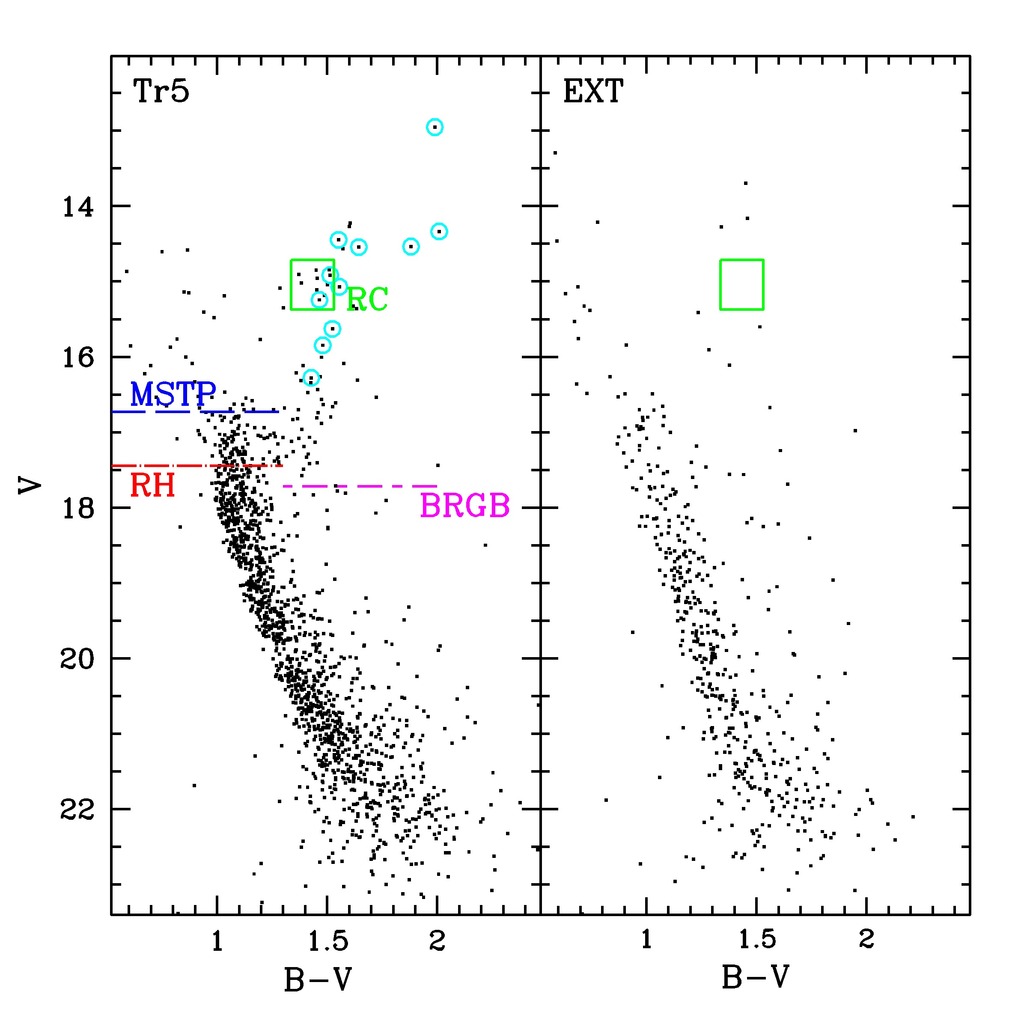}
\caption{Left panel: CMD for stars inside 3 arcmin from the cluster centre; the open circles are the RV candidate members  (from literature, see Table~\ref{tabcat})) together with our spectroscopic targets (see next sections) located inside this radius. Right panel: external field of the same area. We highlight the RC locus in both CMDs and the main evolutionary phases of the cluster in the left panel.}
\label{figext}
\end{figure}

\begin{table*}
\centering
\label{tabcat}
\caption{Stars in common between our photometry and CaT spectroscopy (RV$_{04}$ from \citealt{cole04}, RV$_{07}$ from \citealt{carrera07}) and FLAMES spectra (RV$_{UV}$, \citealt{monaco}). } 
\setlength{\tabcolsep}{1.5mm}
\begin{tabular}{crccccrrrl}
\hline 
ID     & Other    & RA     & Dec     & B & V & RV$_{04}$     & RV$_{07}$      & RV$_{UV}$	   & Notes\\
       &(WEBDA)   &(h:m:s) &(d:p:s)  &   &   &(km~s$^{-1}$)  &(km~s$^{-1}$)   &(km~s$^{-1}$)  &	 \\
\hline
 100016 & 1318 & 6:36:53.47 & +09:25:34.6 & 16.495 & 15.099 &  -   &   -   & 48.1 &   \\
 200008 & 1063 & 6:36:24.23 & +09:25:15.3 & 16.418 & 14.537 &  -   &  51.8 &  -   &   \\
 200015 & 4791 & 6:36:33.15 & +09:33:03.7 & 16.500 & 15.002 &  -   &   -   & 50.0 &   \\
 200027 & 2565 & 6:36:28.48 & +09:28:13.4 & 17.326 & 15.846 & 60.5 &   -   &  -   &   \\
 200070 & 1834 & 6:36:35.56 & +09:26:48.0 & 17.706 & 16.278 & 50.8 &   -   &  -   &   \\
 200113 &  488 & 6:36:40.41 & +09:23:45.1 & 18.294 & 16.557 & 57.1 &   -   &  -   &   \\
 204788 &  833 & 6:36:42.07 & +09:24:34.2 & 16.696 & 14.998 & 54.4 &  39.8 &  -   &   \\
 204791 & 3416 & 6:36:40.20 & +09:29:47.8 & 16.521 & 15.062 &  -   &   -   & 49.8 &   \\
 204801 & 2280 & 6:36:35.99 & +09:27:35.2 & 16.709 & 15.244 & 47.8 &  35.9 &  -   &   \\
 204816 & 1401 & 6:36:28.10 & +09:25:56.3 & 16.650 & 15.349 & 14.8 &   -   &  -   & NM\\
 204823 & 2324 & 6:36:24.83 & +09:27:46.6 & 16.627 & 15.072 & 53.1 &  41.6 &  -   &   \\
 204826 & 2579 & 6:36:23.93 & +09:28:17.2 & 17.153 & 15.629 & 47.8 &  40.9 &  -   &   \\
 204847 & 5099 & 6:36:18.81 & +09:34:06.6 & 14.459 & 12.352 &  -   &  48.7 &  -   &   \\
 204849 & 4219 & 6:36:23.44 & +09:31:37.7 & 14.857 & 12.598 &  -   &  31.4 &  -   &   \\
 204851 & 1935 & 6:36:33.12 & +09:27:00.9 & 14.941 & 12.952 & 53.5 &  45.9 &  -   &   \\
 204856 & 1378 & 6:36:19.26 & +09:25:58.7 & 15.360 & 13.414 &  -   &  42.9 &  -   &   \\
 204859 & 4811 & 6:36:47.32 & +09:32:59.6 & 15.421 & 13.734 &  -   &  54.0 &  -   &   \\
 204864 & 1214 & 6:36:42.29 & +09:25:25.7 & 15.704 & 14.014 & 55.7 &  34.4 &  -   &   \\
 204871 & 1305 & 6:36:24.54 & +09:25:46.7 & 16.345 & 14.336 & 64.9 &  61.0 &  -   &   \\
 204873 & 3066 & 6:36:36.47 & +09:29:08.1 & 16.000 & 14.448 & 50.7 &  37.7 &  -   &   \\
 204876 & 3354 & 6:36:29.31 & +09:29:45.2 & 16.188 & 14.544 & 53.7 &  29.2 &  -   &   \\
 204877 & 3763 & 6:36:34.11 & +09:30:34.6 & 16.138 & 14.566 & -3.9 & -20.6 &  -   & NM\\
 204885 & 1026 & 6:36:42.96 & +09:25:02.3 & 16.428 & 14.823 &  -   &  29.7 &  -   &   \\
 204896 & 4649 & 6:36:48.04 & +09:32:33.5 & 16.602 & 14.999 &  -   &   -   & 47.3 & NM\\
   -	& 6223 & 6:36:38.6  & +09:38:52.6 & 16.57  & 15.08  &  -   &   -   & 50.8 &   \\
% 204783 &   -  & 6:36:44.06 & +09:29:23.5 & 16.690 & 15.223 &  -   &   -   &  -   & 48.1 &   \\
% 204892 &   -  & 6:36:31.24 & +09:28:10.9 & 16.431 & 14.917 &  -   &   -   &  -   & 51.8 &   \\
% 204893 &   -  & 6:36:41.94 & +09:28:11.7 & 16.316 & 14.933 &  -   &   -   &  -   & 49.2 &   \\
\hline
\end{tabular}
\end{table*}

\subsection{Cluster centre and CMD}\label{sec:CMD}
Exploiting the deep and precise photometry obtained with WFI and its large FoV, we re-determined the centre of the cluster following the approach described in \cite{donati12}. Briefly, we selected the densest region on the images by looking for the smallest coordinates interval that contains 70\% of all the stars. The centre is obtained as the average right ascension and declination when the selection is iterated twice. 
For a more robust estimate, several magnitude cuts have been considered and the corresponding results averaged.  The root mean square (r.m.s.) on the centre coordinates is about 2$\arcsec$. We found
RA(J2000)=06:36:28.22, %(hour$:\arcmin:\arcsec$), 
Dec(J2000)=+09:28:04.26, %($\degr:\arcmin:\arcsec$), 
significantly different from the values cited in WEBDA (RA=06:36:42, Dec=+09:26:00 both referred to J2000).
From the density profile it was also possible to define the apparent diameter of the cluster. We estimated $d=26\arcmin\pm4\arcmin$ using the radius at which the density profile flattens at the background density value; the cluster
appears larger than found by P04, probably thanks to the advantage of a wider FoV.

\begin{figure*}
\includegraphics[scale=1.5]{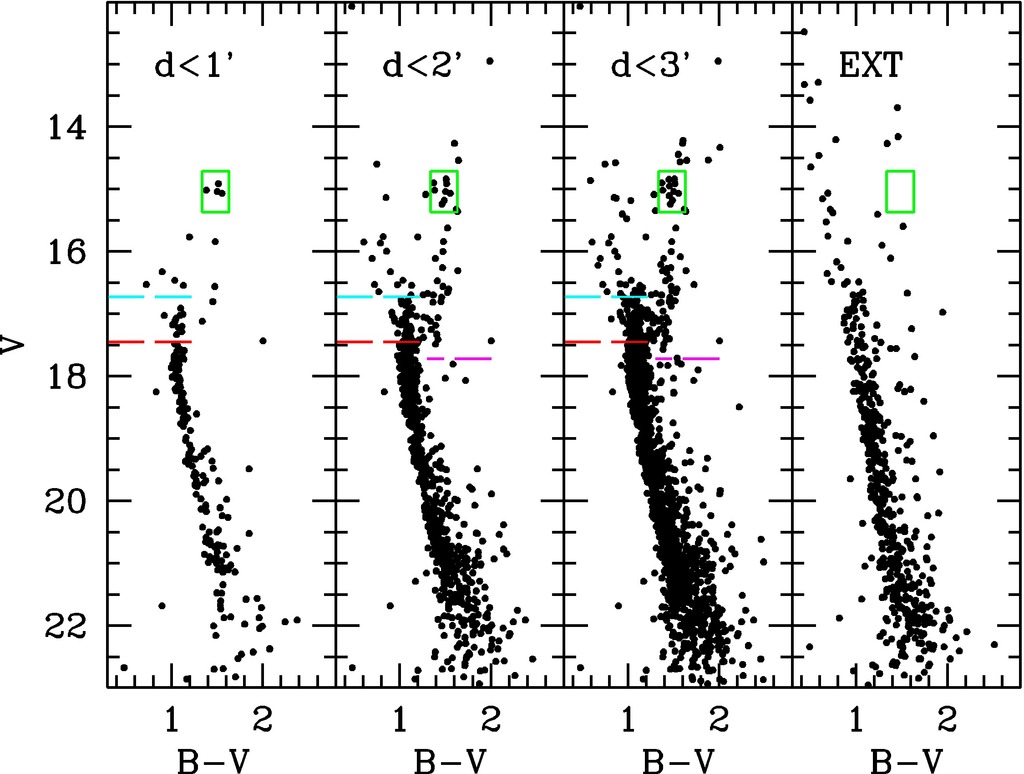}
\caption{CMDs  for different distances from the cluster centre (first three panels from left to right: d$<1\arcmin$, d$<2\arcmin$, and d$<3\arcmin$), compared with an external field of a circular area of  $3\arcmin$ radius. We highlight the main evolutionary phases: RH, MSTP, and RC. For a better comparison the RC box is also shown in the CMD of the external field.}
\label{figrad}
\end{figure*}

With this information we analysed the morphology of the cluster CMD.
As apparent from Fig.~\ref{figext}, where we show Tr~5 and the comparison field, the cluster is populous. The MS is well visible, as is the rich RC. The high contamination by field objects clearly complicates the interpretation of the evolutionary features, especially the red-hook (RH, the reddest point on the MS before the overall contraction), the MS termination point (MSTP, the most luminous level of the MS phase before the runaway to the red), and the RGB. By considering only the innermost part of the cluster it is possible to minimise the pollution of field interlopers. Furthermore, we can use information on 
membership from spectroscopy. Table~4 lists the stars in 
common in our photometry and published spectroscopic studies, which are also shown in Fig.~\ref{figext} (left-hand panel) as large open circles. In Fig.~\ref{figrad} we show the CMDs of Tr~5 for different distances from the cluster centre (from left to right: 1$\arcmin$, 2$\arcmin$, 3$\arcmin$). 
We used the radial plots to identify the evolutionary features of the cluster employed in the following analysis (see Sec.~\ref{CMDsynth}), and we identified:
\begin{itemize}
  \item the RC at $V\sim15$ mag and $(B-V)\sim1.55$ mag;
  \item an extended RGB, with the base of the RGB (BRGB) at $V\sim17.4$ mag;
  \item the MSTP at $V\sim16.8$ mag and the blue edge (BE) of the MS at $(B-V)\sim1$ mag;
  \item the RH at $V\sim17.3$ mag;
  \item and the MS extending down to $V\sim23$ mag.
\end{itemize}

\subsection{Differential reddening}\label{DR}
The CMD of Tr~5 (see Figs.~\ref{figext}, \ref{figrad}) shows a MS which has a non negligible extension in colour. Such broadening cannot be explained only by the photometric error, which is small especially at bright magnitudes. 
Apart from contamination by field stars, other possible influencing agents are the presence of binary systems, a spread in age or in metallicity, and differential reddening. 
We exclude a large spread in age, which would not explain the evident broadening of the lower MS, or in metallicity, since no open cluster has ever convincingly been demonstrated to have it. Binarism, instead, has always been found in OCs, and indeed we do take it into account in the CMD analysis, but it leads to a smaller colour spread with a typical inclination with respect to the single
stars MS. 
Differential reddening (DR) seems the most plausible explanation because the patchy structure of dust intercepted along the line of sight has the net effect to shift the colour and magnitude of the stars along the reddening vector: the denser the ISM the redder and fainter the observed colour and magnitude. This condition is particularly likely for objects residing on the disc and with a high average reddening,  as is the case of Tr~5 ($b\sim1^o$, $E(B-V)\sim0.6$). Furthermore, Tr~5 lies in the vicinity of the star forming region NGC~2264 and of the Cone nebula.
Tr~5 is not a peculiar case. In fact, non negligible DR has been found in many other clusters \cite[see e.g.,][only to name a few cases]{platais,brogaard,donati14a,donati14b}.

\begin{figure}                      
\includegraphics[scale=0.9]{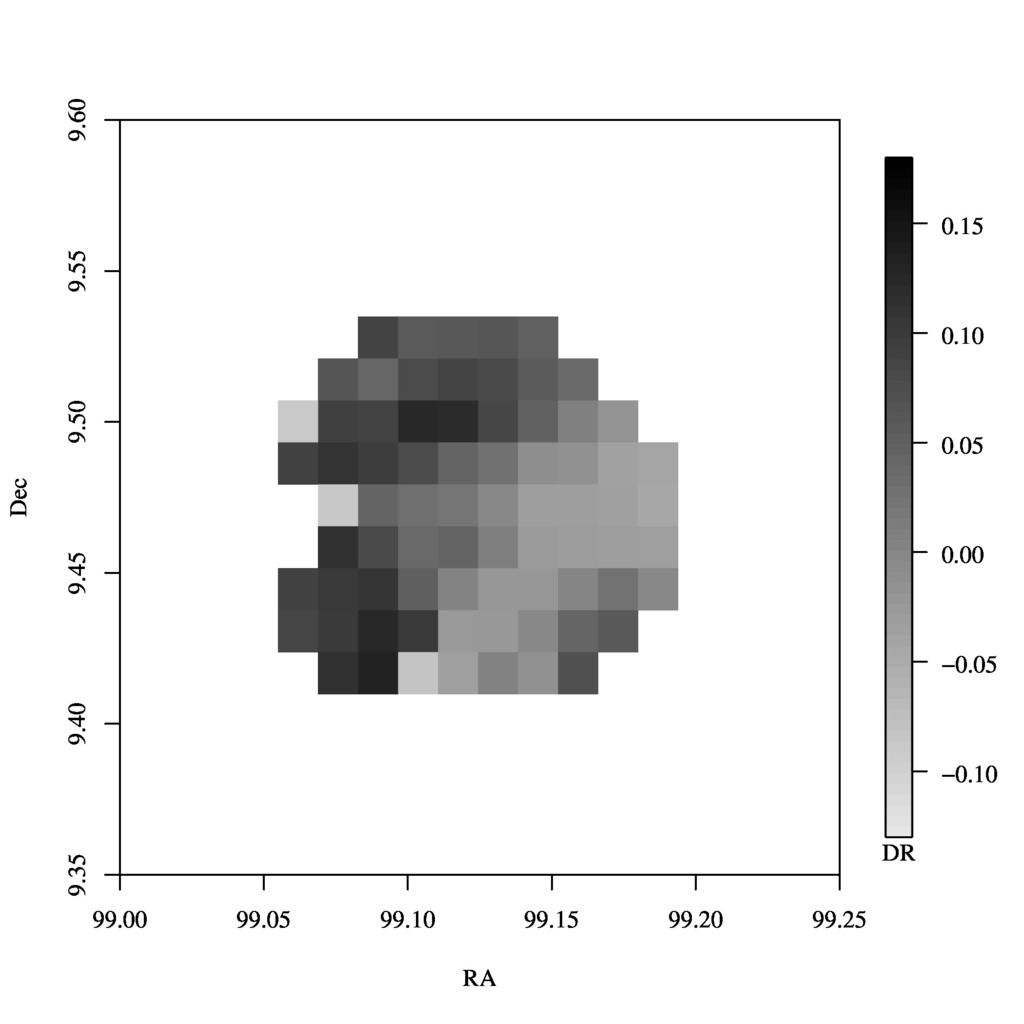}
\includegraphics[scale=0.9]{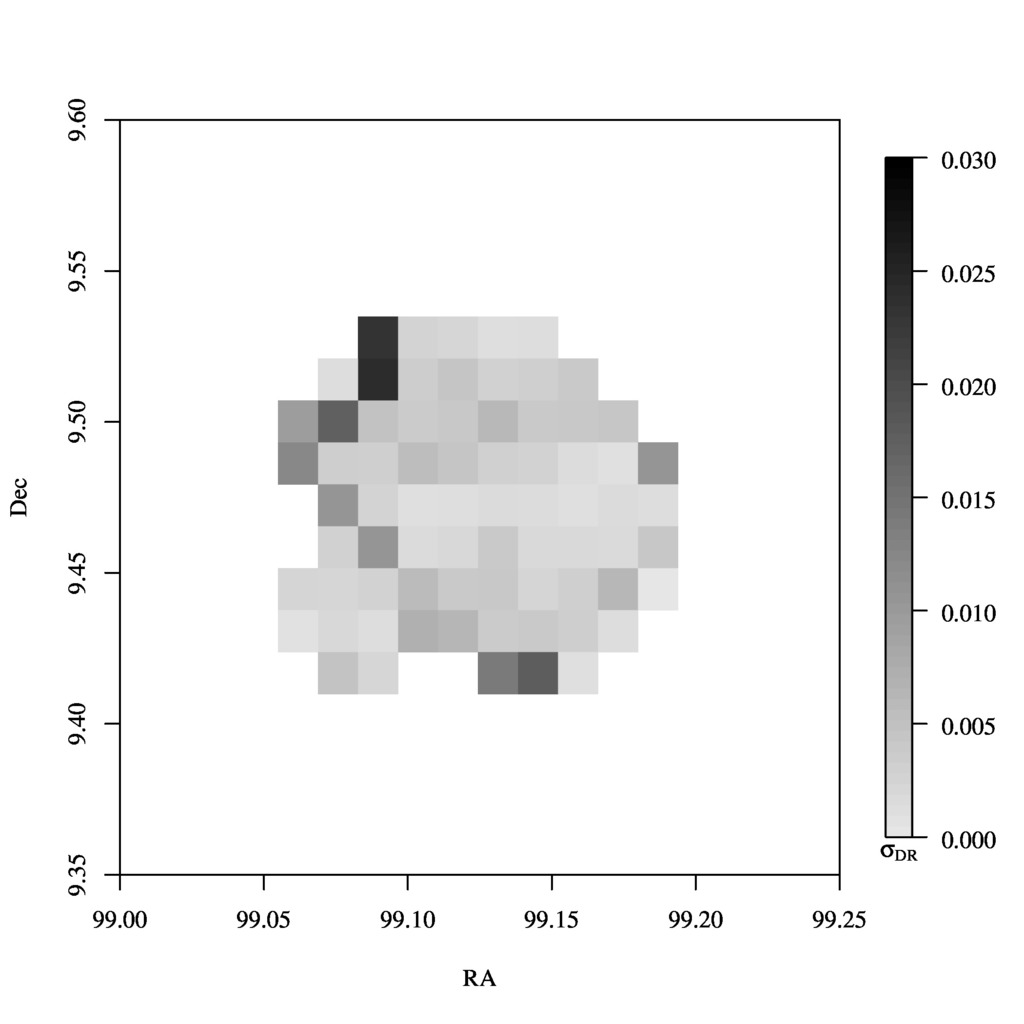}
\caption{These plots show the reddening map (upper panel) and the corresponding error (lower panel) in RA and Dec (expressed in degrees) obtained for Tr~5. The spatial resolution used is $50\arcsec\times50\arcsec$. The colour code indicates where DR (upper panel) and the error (lower panel) are stronger (dark colour) or weaker (light colour).}
\label{fig:drgrid}
\end{figure}

To quantify the effect of DR we use a method based on the one described by \cite{mil12}, but adapted to the case of OCs. 
As explained in more details in \cite{donati14a,donati14b}, we use stars selected within a region on the MS and in small spatial areas (here, $50\arcsec\times50\arcsec$) and compute their average distance along the reddening vector from a fiducial line (i.e. their DR).
We evaluated the DR only inside 4$\arcmin$ from the cluster centre, where the density of stars is about 50\% higher than in the external region of our FoV. We avoided the external parts, where our estimation of DR could be severely jeopardised by the presence of too many field interlopers. 
The resulting DR map and the map of the error on the average are shown in Fig.~\ref{fig:drgrid}.

\begin{figure}
\includegraphics[scale=0.9]{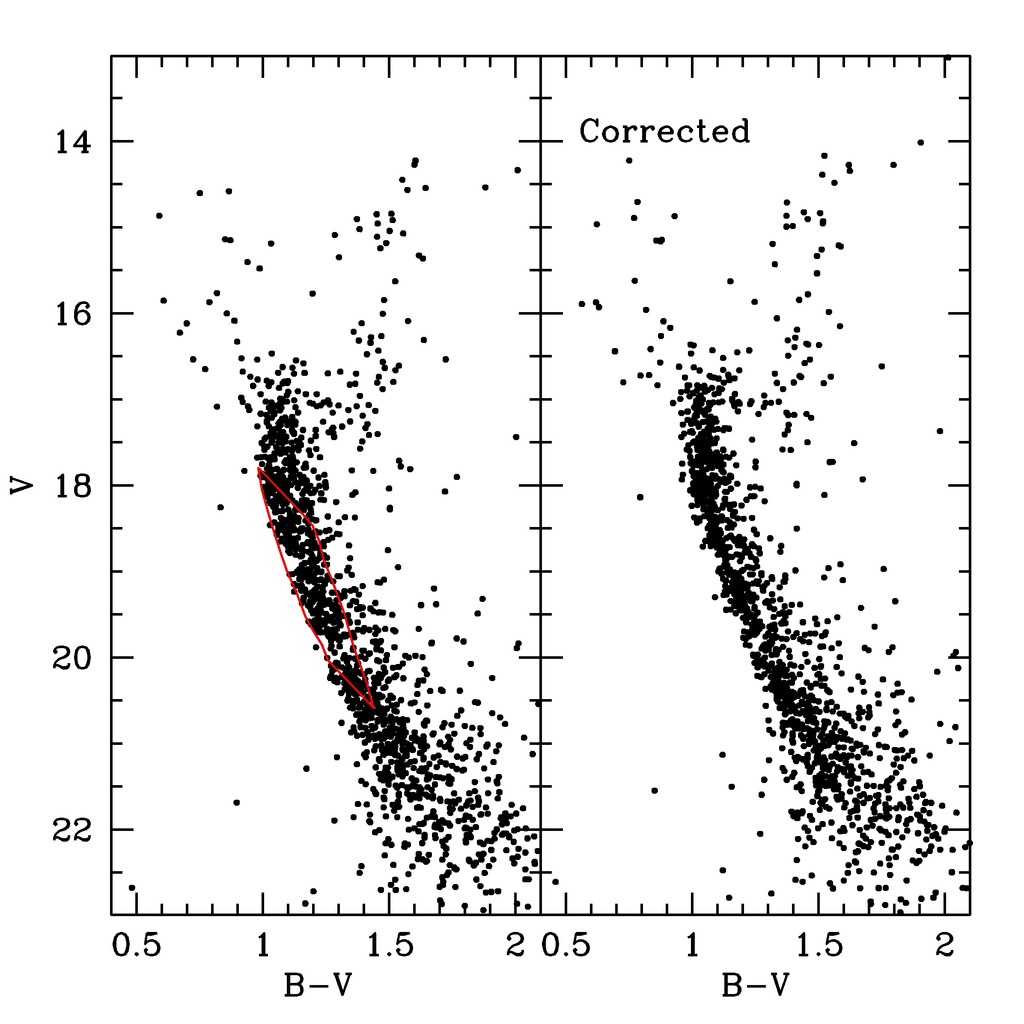}
\caption{On the left: the observed CMD within 3 arcmin, with the 
red curve enclosing the MS portion used to estimate the DR effect. On the right: the CMD on the same region after the correction for the resulting DR.}
\label{fig:drbv}
\end{figure}

The DR ranges between about $-0.1$ and $+0.15$. The standard deviations of the DR measurements in each cell of the grid range from less than 0.01 mag up to 0.07 mag, with an average of 0.02 mag. This strong variation within the same spatial cell of the grid are in part due to the intrinsic variation of the reddening on the cluster face coupled with the strong contamination of field interlopers. The combination of these two factors limits the precision of the DR measurements and translates into errors on the average values within each cell up to 0.03 mag in the worst case. In Fig.~\ref{fig:drbv} we show the comparison between the CMDs with and without
the correction for DR. The MS appears considerably improved, with a much
tighter extension in colour; also the giant phases seem to be much better
defined, in particular the SGB, RGB, and the RC. The sequence of binaries is much more evident after the correction and it still contributes to the spread of the MS. Their impact on the CMD appearance is quantified using the synthetic CMD method (see Section \ref{CMDsynth}). 

\subsection[]{Spectroscopy observations and reductions}\label{obsspec}
The spectra of three RC stars were acquired in service mode at the ESO Very
Large Telescope (VLT) with the high-resolution spectrograph UVES (Ultraviolet
VLT Echelle Spectrograph, \citealt{dekker}). None of them is in common with the samples
of \cite{cole04}, \cite{carrera07}, and \cite{monaco} but all three are cluster
members, according to their radial velocity (RV, see below). Details of the
observations are given in Table~\ref{tabspe1}, where airmass and seeing are
approximate average values for the three spectra in each night. We employed a
slit width of 1.2~arcsec ($R\simeq40000$) and the dichroic, that splits the
light into a blue channel (CD\#2, $\lambda\lambda$ 3280-4560 \AA) and a red
channel (CD\#3, $\lambda\lambda$ 4720-5800, 5810-6830 \AA). 

We used the spectra reduced (extracted, wavelength calibrated, and sky
subtracted) by ESO and retrieved from the ESO Advanced Data Products
archive. Individual spectra have signal-to-noise (S/N) less than 10 in the blue
chip, about 25 in the lower chip of the red channel, and about 35 in the upper chip.
We discarded the blue spectra and averaged (using the median) the individual red
spectra for each star using {\sc iraf}: the S/N of the combined spectra in the
lower and upper red channels are reported in Table~\ref{tabspe2}.

\begin{table}
\centering
\caption{Log of the UVES observations\label{tabspe1}}
\setlength{\tabcolsep}{1.5mm}
\begin{tabular}{cccccrc}
\hline
ID   & Other    & Date obs   & exptime & seeing  &airmass \\
     &(WEBDA)   &(UT)        &(s)      &(arcsec) &        \\
       \hline
204893 &2604 & 2005-01-04 & 3$\times$860 &2.8  & 1.50\\
       &     & 2005-01-05 & 3$\times$860 &1.2  & 1.23\\
204783 &3236 & 2005-01-08 & 3$\times$960 &0.9  & 1.21\\
204892 &2553 & 2005-02-12 & 3$\times$860 &1.0  &1.23 \\
\hline
\end{tabular}
\end{table}

We measured the equivalent widths (EW) of atomic lines in the combined spectra
using {\sc doo}p \citep{doop}, an automated wrapper for {\sc daospec}
\citep{daospec}, originally devised for the analysis of the many thousands of
spectra obtained by the Gaia-ESO Survey \citep[see][for a description]{gilmore,randich}. The line list employed
was prepared for the Gaia-ESO Survey (by the line list group; a paper by Heiter et al. is in
preparation) and is based on VALD3 \citep{vald3} atomic data. The measured EWs
and their errors, the excitation potentials and $\log gf$ values of the transitions,
and the {\sc daospec} quality parameter for each line
can be found in Table~\ref{ew} (available in its entirety only in electronic form).

The observed RV is an output of DAOSPEC\footnote{As shown in Table~\ref{tabspe1}, the spectra of each star were taken consecutively, hence RVs were measured on the averaged spectra given that there were no significant shifts in wavelength from spectrum to spectrum.}, and telluric absorption features were
used to correct for any wavelength calibration shifts or misalignment's within
the slit, following the procedure described by \citet{pancino10}. Heliocentric
corrections were applied and the resulting heliocentric RVs are
listed in Table~\ref{tabspe2}, together with coordinates, magnitudes, and
atmospheric parameters (see Section~\ref{abu}).

\begin{table*}
\centering
\caption{Photometry, coordinates at J2000.0, RV,  S/N in the lower and upper CCDs, and atmospheric parameters of the three stars
observed with UVES.\label{tabspe2}}
\setlength{\tabcolsep}{1.5mm}
\begin{tabular}{ccccccccccccc}
\hline
ID              & B  & V & K & RA  & Dec & RV  & S/N & S/N & $T_{eff}$ &$\log g$ &$v_t$ & $[$Fe/H$]$ \\
      & & &(2MASS) &(h:m:s) &(d:p:s) &(km~s$^{-1})$  & (lower) & (upper) &(K) & &(km~s$^{-1}$) & \\
\hline
204892  & 16.431 & 14.917 &10.588 & 6:36:31.24 & +09:28:10.9 & 51.8$\pm$1.1 &30 &55 &4750 &2.60 &1.40 &--0.40\\
204893  & 16.316 & 14.933 &11.018 & 6:36:41.94 & +09:28:11.7 & 49.2$\pm$1.0 &45 &70 &5000 &2.70 &1.70 &--0.41\\
204783  & 16.690 & 15.223 &11.134 & 6:36:44.06 & +09:29:23.5 & 48.1$\pm$0.9 &35 &60 &4850 &2.80 &1.30 &--0.40\\
\hline
\end{tabular}
\end{table*}

\section{Abundance analysis}\label{abu}
Abundances and atmospheric parameters were derived with {\sc gala} 
\citep{gala}, an
automated wrapper for the Kurucz abundance calculation code
\citep{kurucz,sbordone}, employing Atlas atmospheric 
models\footnote{{\sc gala} allows for a quick computation with both the Atlas
and the MARCS \citep{marcs} atmospheric models. The use of MARCS models provides
almost identical  results: $<\Delta\rm{T}_{\rm{eff}}>=$+17~K, $<\Delta\log
g>=-0.1$~dex, $<\Delta v_t>=-0.03~$km~s$^{-1}$, $<\Delta$[Fe/H]$>=$+0.01~dex,
with all differences computed in the sense MARCS minus Atlas.} \citep[based on
the grid by][]{castelli}. 
{\sc gala} uses the classical EW method, where initial first  guesses of the
atmospheric parameters are refined by erasing trends of Fe abundance  with the
excitation potential (to refine T$_{\rm{eff}}$), with EW (to refine  v$_t$), and
with wavelength (as a general sanity check of the method). Surface  gravity is
refined by imposing ionisation balance, i.e., that Fe~{\sc i} and Fe~{\sc ii}
give  the same Fe abundance, within the uncertainties. Initial guesses for the
atmospheric parameters were obtained from stellar models and our experience
with  sub-solar metallicity RC stars (T$_{\rm{eff}}\simeq$5000~K, $\log g\simeq
2.5$~dex, $v_t\simeq 1$~km~s$^{-1}$) and {\sc gala} was let free to seek for
convergence  starting from there: the final, adopted parameters are listed in 
Table~\ref{tabspe2}. We found an average metallicity [Fe/H]=$-0.40$~dex, based
on $\simeq$170  Fe~{\sc i} lines and $\simeq$20 Fe~{\sc ii} lines, confirming
that Tr~5 is a metal-poor open  cluster. Its position in the Galactic disc
metallicity distribution will be discussed in Sect.~\ref{summary}.

We also measured abundances for the light elements Al (based on two or three
lines, depending on the star) and Na (3 lines); the $\alpha$-elements Ca
($\simeq$9 lines), Mg (3 lines), Si ($\simeq$22 lines), and Ti  ($\simeq$50
Ti~{\sc i} and $\simeq$10 Ti~{\sc ii} lines); the iron-peak elements Sc
($\simeq$3 Sc~{\sc i} and $\simeq$10 Sc~{\sc ii} lines), V ($\simeq$25 lines),
Cr ($\simeq$22 lines), Co ($\simeq$20 lines), and Ni ($\simeq$53 lines); and the
heavy elements La (2 lines), Pr (2 lines), Nd ($\simeq$19 lines), and Y
($\simeq$7 lines).  All our abundances were derived in LTE (Local Thermodynamic
Equilibrium) and without corrections for HFS (Hyper Fine Structure) effects.
Abundances of the  species listed in Table~\ref{tababu} were derived as the
weighted average of the abundances provided by the single lines, with their
error, computed as the sigma on 
the weighted average, divided by the square root of the number of used lines. The
adopted solar composition  was the one by \citet{grevesse96}.

The  uncertainty owing to the continuum placement was between 2\% and 5\%
judging from the {\sc daospec} residuals after removal of the fitted lines;
globally this corresponds to approximately an error of 5-10~m\AA\ on the single
lines, that can amount to approximately 0.05~dex at most in the [Fe/H]
abundance, for example. The additional uncertainty implied by the choice of
atmospheric parameters was estimated using the \citet{cayrel} method, which
takes automatically  into account correlations among the parameters by altering
just one of them (in our case T$_{\rm{eff}}$, by $\pm$100~K) and keeping it
fixed to a wrong value, while re-optimising the other parameters. The resulting
uncertainties are  reported in Table~\ref{tababu} between parenthesis.

\begin{table}
\begin{textbf}
\centering
\caption{Equivalent widths of the used atomic species. Only the first few lines are reproduced here, the complete table can be found in its electronic form at CDS.\label{ew} }
\setlength{\tabcolsep}{1.5mm}
\begin{tabular}{cccccccc}
\hline
ID   & $\lambda$ & Species & $\chi_{\rm{ex}}$ & log$gf$  & EW     & $\delta$EW & Q \\
     &(\AA)      &         & (eV)             & (dex)   & (m\AA) & (m\AA)     &   \\
       \hline
204893 & 4808.148 & FeI &  3.250 &  -2.690 &  45.60 &  2.66 &  0.837 \\
204893 & 4809.938 & FeI &  3.570 &  -2.620 &  22.50 &  2.04 &  0.868 \\
204893 & 4873.751 & FeI &  3.300 &  -2.960 &  28.90 &  3.18 &  1.340 \\
204893 & 4874.353 & FeI &  3.070 &  -3.088 &  36.10 &  2.56 &  0.938 \\
204893 & 4896.439 & FeI &  3.880 &  -1.950 &  45.40 &  1.76 &  0.522 \\
\hline
\end{tabular}
\end{textbf}
\end{table}

Most abundance ratios appear solar within the uncertainties, although  Mg, Al,
Pr, Nd, and La are slightly supersolar as seen in other OCs of similar
metallicity \citep{bragaglia08,sestito08,pancino10,carrera11,yong12}.  The three
stars share a very homogeneous composition in all the examined species; this
conclusion is robust even if we did not correct for departure from LTE  or HFS,
since they are in the same evolutionary status and their atmospheric  parameters
are essentially identical.

Very similar conclusions can be reached using the results published in
\cite{monaco}. That paper is devoted to the analysis of a Li-rich RC  star and
the properties of the normal stars are not discussed, so we cannot make here a
detailed comparison. However, from the tables in their Appendix we see  that the
four stars that they consider member of Tr~5 (all in the RC phase) have 
properties similar to the ones we derived for our three stars. In particular, we
find $<RV>=49.7$ (rms 1.9) km~s$^{-1}$, while they have $<RV>=49.8$ (rms 1.1)
km~s$^{-1}$, the average metallicities are $-0.40$ and $-0.49$ dex, 
respectively, and we also have similar abundances for the elements in common.
Finally,  also in their analysis the stars show a very homogeneous composition,
without any anomalous spread. 

\begin{table*}
\centering
\caption{Abundance ratios (with respect to neutral iron), with internal errors 
($\sigma/\sqrt{(nlines)}$) and sensitivity of each abundance to changes in 
atmospheric parameters between parenthesis (see text). The cluster average and sigma are
also indicated, along with the Solar reference abundance (see text). 
\label{tababu}}
\setlength{\tabcolsep}{1.5mm}
\begin{tabular}{lcccccccccccccc}
\hline
Ratios & star 204892 & star 204893 & star 204783 & Tr~5 & Sun\\
& (dex) & (dex) & (dex) & (dex) & \\
\hline
$[$Fe~I/H$]$ & --0.40 $\pm$ 0.01 ($\pm$ 0.06) & --0.41 $\pm$ 0.01 ($\pm$ 
0.06) & --0.40 $\pm$ 0.01 ($\pm$ 0.09) & --0.40 $\pm$ 0.01 & 7.50 \\
$[$Fe~II/H$]$ & --0.44 $\pm$ 0.01 ($\pm$ 0.09) & --0.42 $\pm$ 0.02 
($\pm$ 0.03) & --0.36 $\pm$ 0.02 ($\pm$ 0.08) & --0.41 $\pm$ 0.04 & 7.50 \\
\hline
$[$Na~I/Fe$]$ & +0.00 $\pm$ 0.03 ($\pm$ 0.06) & --0.02 $\pm$ 0.06 ($\pm$ 
0.06) & +0.05 $\pm$ 0.05 ($\pm$ 0.08) & +0.01 $\pm$ 0.04 & 6.33 \\
$[$Mg~I/Fe$]$ & +0.17 $\pm$ 0.02 ($\pm$ 0.03) & +0.20 $\pm$ 0.03 ($\pm$ 
0.06) & +0.24 $\pm$ 0.03 ($\pm$ 0.06) & +0.20 $\pm$ 0.04 & 7.58 \\
$[$Al~I/Fe$]$ & +0.21 $\pm$ 0.07 ($\pm$ 0.06) & +0.20 $\pm$ 0.04 ($\pm$ 
0.06) & +0.25 $\pm$ 0.03 ($\pm$ 0.08) & +0.22 $\pm$ 0.03 & 6.47 \\
$[$Si~I/Fe$]$ & --0.02 $\pm$ 0.02 ($\pm$ 0.03) & +0.01 $\pm$ 0.03 ($\pm$ 
0.02) & +0.05 $\pm$ 0.03 ($\pm$ 0.03) & +0.01 $\pm$ 0.04 & 7.55 \\
$[$Ca~I/Fe$]$ & +0.05 $\pm$ 0.03 ($\pm$ 0.09) & +0.05 $\pm$ 0.02 ($\pm$ 
0.08) & +0.09 $\pm$ 0.02 ($\pm$ 0.12) & +0.06 $\pm$ 0.02 & 6.36 \\
$[$Sc~I/Fe$]$ & --0.11 $\pm$ 0.03 ($\pm$ 0.15) & --0.11 $\pm$ 0.07 
($\pm$ 0.18) & --0.01 $\pm$ 0.03 ($\pm$ 0.15) & --0.08 $\pm$ 0.06 & 3.17 \\
$[$Sc~II/Fe$]$ & +0.18 $\pm$ 0.04 ($\pm$ 0.06) & +0.20 $\pm$ 0.04 ($\pm$ 
0.08) & +0.15 $\pm$ 0.04 ($\pm$ 0.06) & +0.18 $\pm$ 0.03 & 3.17 \\
$[$Ti~I/Fe$]$ & --0.04 $\pm$ 0.01 ($\pm$ 0.12) & --0.05 $\pm$ 0.01 
($\pm$ 0.14) & +0.02 $\pm$ 0.01 ($\pm$ 0.15) & --0.02 $\pm$ 0.04 & 5.02 \\
$[$Ti~II/Fe$]$ & +0.12 $\pm$ 0.03 ($\pm$ 0.08) & +0.04 $\pm$ 0.04 ($\pm$ 
0.06) & +0.08 $\pm$ 0.02 ($\pm$ 0.06) & +0.08 $\pm$ 0.04 & 5.02 \\
$[$V~I/Fe$]$ & +0.06 $\pm$ 0.02 ($\pm$ 0.15) & --0.03 $\pm$ 0.02 ($\pm$ 
0.17) & +0.08 $\pm$ 0.03 ($\pm$ 0.18) & +0.04 $\pm$ 0.06 & 4.00 \\
$[$Cr~I/Fe$]$ & --0.08 $\pm$ 0.03 ($\pm$ 0.09) & +0.04 $\pm$ 0.05 ($\pm$ 
0.11) & --0.09 $\pm$ 0.03 ($\pm$ 0.12) & --0.04 $\pm$ 0.07 & 5.67 \\
$[$Co~I/Fe$]$ & +0.11 $\pm$ 0.03 ($\pm$ 0.09) & +0.03 $\pm$ 0.03 ($\pm$ 
0.14) & +0.10 $\pm$ 0.03 ($\pm$ 0.11) & +0.08 $\pm$ 0.04 & 4.92 \\
$[$Ni~I/Fe$]$ & --0.07 $\pm$ 0.02 ($\pm$ 0.05) & --0.06 $\pm$ 0.02 
($\pm$ 0.06) & --0.05 $\pm$ 0.02 ($\pm$ 0.08) & --0.06 $\pm$ 0.01 & 6.25 \\
$[$Y~II/Fe$]$ & --0.08 $\pm$ 0.04 ($\pm$ 0.06) & --0.07 $\pm$ 0.04 
($\pm$ 0.09) & +0.05 $\pm$ 0.09 ($\pm$ 0.06) & --0.03 $\pm$ 0.07 & 2.24 \\
$[$La~II/Fe$]$ & +0.16 $\pm$ 0.07 ($\pm$ 0.09) & +0.10 $\pm$ 0.06 ($\pm$ 
0.14) & +0.22 $\pm$ 0.02 ($\pm$ 0.09) & +0.16 $\pm$ 0.06 & 1.17 \\
$[$Pr~II/Fe$]$ & +0.29 $\pm$ 0.17 ($\pm$ 0.08) & +0.03 $\pm$ 0.12 ($\pm$ 
0.14) & +0.15 $\pm$ 0.12 ($\pm$ 0.09) & +0.16 $\pm$ 0.13 & 0.71 \\
$[$Nd~II/Fe$]$ & +0.21 $\pm$ 0.03 ($\pm$ 0.08) & +0.19 $\pm$ 0.03 ($\pm$ 
0.12) & +0.19 $\pm$ 0.03 ($\pm$ 0.09) & +0.20 $\pm$ 0.01 & 1.50 \\
\hline
\end{tabular}
\end{table*}

\section[]{Clusters parameters using synthetic colour-magnitude diagrams}\label{CMDsynth}
The age, distance modulus, metallicity, reddening, differential reddening, and
binary fraction of the cluster are estimated using the synthetic colour
magnitude diagram technique \citep[see][]{tosi91} as done in all the papers of the BOCCE project (see e.g., \citealt{cignoni11,donati12,donati14a} and references therein). 
For a detailed description, see \cite{bt06}. Briefly, we compute a grid of synthetic CMDs in the age-metallicity-distance-reddening space
using three sets of stellar evolution models, i.e. the Padova \citep{bbc93,fag94},
the FRANEC \citep{franec}, and the FST \citep{fst} tracks, chosen to test the effect of different input physics on the derivation of
cluster fundamental parameters. In fact, these models use different prescriptions
for the treatment of convection, going from no overshooting (FRANEC), to the 
standard description of overshooting through parametrisation of the mixing length (Padova), to overshooting treated using the so-called full spectrum of turbulence modellisation (FST). The last models have also the possibility of
choosing between three different levels of overshooting (none, moderate, and high). These tracks offer only a few possible metallicities, and we do not
interpolate between them. In particular, the metallicity values we 
normally use for
OCs are $Z=$0.02, 0.008, 0.004 (Padova); $Z=$0.02, 0.01, 0.006 (FRANEC); and
$Z=$0.02, 0.006 (FST).
All these models are available only with one choice of helium and solar-scaled $\alpha$ elements. The lack of a fine
metallicity grid and of different levels of $\alpha$-enhancement prevents
a direct comparison with the metallicity and chemical composition derived
from the spectra.
Although more modern tracks are available nowadays, we favour homogeneity of treatment and continue to use the same ones adopted throughout the BOCCE project.

For Tr~5 all synthetic CMDs are built assuming an approximately instantaneous star formation burst (5 Myr long) and using a single slope Salpeter IMF over the range
in mass covered by the tracks. The photometric conversions from the theoretical effective temperature-luminosity plane to the empirical colour-magnitude plane are obtained with the same conversion tables \citep[see][]{bessel98} for all the sets of tracks.
Cluster parameters are determined by means of the comparison
of the synthetic CMDs with the observed ones. The best fit solution is
chosen as the one that can best reproduce age-sensitive indicators (highlighted in Sec.~\ref{sec:CMD}): the
RH, the RC, the MSTP, the BRGB, the RGB inclination
and colour, and the RC colour. The latter two were used as secondary age
indicators as colour properties are more affected by theoretical uncertainties,
like colour transformation and super-adiabatic convection, while luminosity
constraints are more reliable. 

Multi-colour photometry has generally proved to be fundamental to obtain the best parameters estimation \citep[see, e.g.,][for a discussion]{ahumada13}, especially metallicity. The best fit solutions must reproduce at the same time the observed CMDs in different colours for appropriate distance modulus, reddening, and age. However, for Tr~5 we knew from literature 
spectroscopy that the cluster is rather metal-poor and we confirmed
that with the UVES spectra. This helped us in restricting the possible range of evolutionary tracks. We used our $B,V$ photometry for the synthetic CMD technique coupled with the P04 $V,I$ photometry to help constraining the photometric metallicity.

We estimated the errors on the cluster parameters considering both the instrumental photometric errors and the uncertainties of the fit analysis, as done in \cite{donati12}. The net effect of the former is an uncertainty on the luminosity level and colour of the  
adopted indicators. This affects mainly the estimate of the mean Galactic reddening and distance modulus, as they are directly defined by matching the level and colour of the upper MS and the RH and MSTP indicators.
We must also consider the dispersion in the results arising from the fit analysis. Tr~5 is heavily contaminated, and the definition of important indicators, such as the RC locus or the MSTP, is more uncertain. Hence, we cannot find a unique solution, but only a restricted range of viable solutions. In practice, we select the best fitting synthetic CMD and then take into account the dispersion of the cluster parameters estimates in the error budget. 
The uncertainties are assumed to be of the form:
$$\sigma^2_{E(B-V)}\sim\sigma^2_{(B-V)}+\sigma^2_{fit}$$
$$\sigma^2_{(m-M)_0}\sim\sigma^2_{V}+R_V^2\sigma^2_{E(B-V)}+\sigma^2_{fit}$$
$$\sigma^2_{age}\sim\sigma^2_{fit}$$
Typical photometric errors are $\sim0.04$ on the reddening and  $\sim0.1$ on the distance modulus (assuming negligible the error on $R_V$). The  error resulting from
the fit analysis depends mainly on the uncertainty on the RC level and on the coarseness of the models grid. It is of the order of $\sim0.02$ for the reddening, ranges between 0.01 and 0.05 for the distance modulus, and is about 0.1 Gyr for the age.

\subsection{CMD}
As discussed in Sec.~\ref{DR}, the cluster's MS is much broader than expected from photometric errors. This is due to two factors: the presence of DR and a significant fraction of unresolved binaries. The former aspect is 
discussed and quantified in Sec.~\ref{DR} and for our simulation we compared the synthetic CMDs with the observed CMD corrected for DR.
For binaries, a rough estimate of their fraction was obtained following the method described in \cite{cignoni11}: we defined two CMD boxes, one which encloses (the bluer) MS stars and the other red-ward of the MS in order to cover the binary sequence. To remove the field contamination we subtracted the contribution of field stars falling inside the same CMD boxes of an equal area of the control field. We performed the same computation on regions of different sizes, eventually ending up with an estimate between 20\%  and 35\%. The dispersion on the estimate is mostly due to the spatial fluctuations across the control field. These fractions are probably underestimated, since we are missing binaries hosting  a low mass star, whose properties are close to those of single stars. We assumed a fraction of 25\% for all the simulations.

Even considering the correction for DR and binaries, we were not able to match the broadening of the MS with the synthetic CMD. We were forced to consider an additional random component of DR of the order of 0.07 mag. This value matches the largest dispersion we found in DR measurements, which, as explained in Sec.~\ref{DR}, are limited in precision due to the large intrinsic spatial variation of the DR across the cluster face and to the contamination of field interlopers. 

The best solution for each set of tracks is the one whose synthetic CMD fits ``most'' of the visible MS shape and the RC and MSTP luminosity levels. In general we found RGB and RC colours slightly redder  in the synthetic CMDs than in the observed one. These evolutionary stages depend
on physical parameters like the metallicity, the age, the helium content as well as on more subtle physical assumptions like
the efficiency of core overshooting and the amount of mass loss during the pre-Helium burning phase, see \citet{castellani}. Moreover, the synthetic colours of the coolest phases are also affected by the uncertainties in the photometric conversions. We found an overall good agreement for the other evolutionary phases. Among the metallicity range allowed by the evolutionary tracks adopted within BOCCE, the best solutions were found for sub-solar metallicity ($Z<0.008$). This is the case in which we were able to obtain a good fit of both $V,B-V$ and $V,V-I$ observational CMDs after adopting the standard extinction law ($E(V-I)=1.25\times E(B-V)$, $R_V=3.1$, see \citet{dean_78}). 
The interval of confidence of the cluster age turned out  to be between 2.9 Gyr and 4.0 Gyr (for models without and with overshooting, respectively).

\begin{figure*}
\includegraphics[scale=1.5]{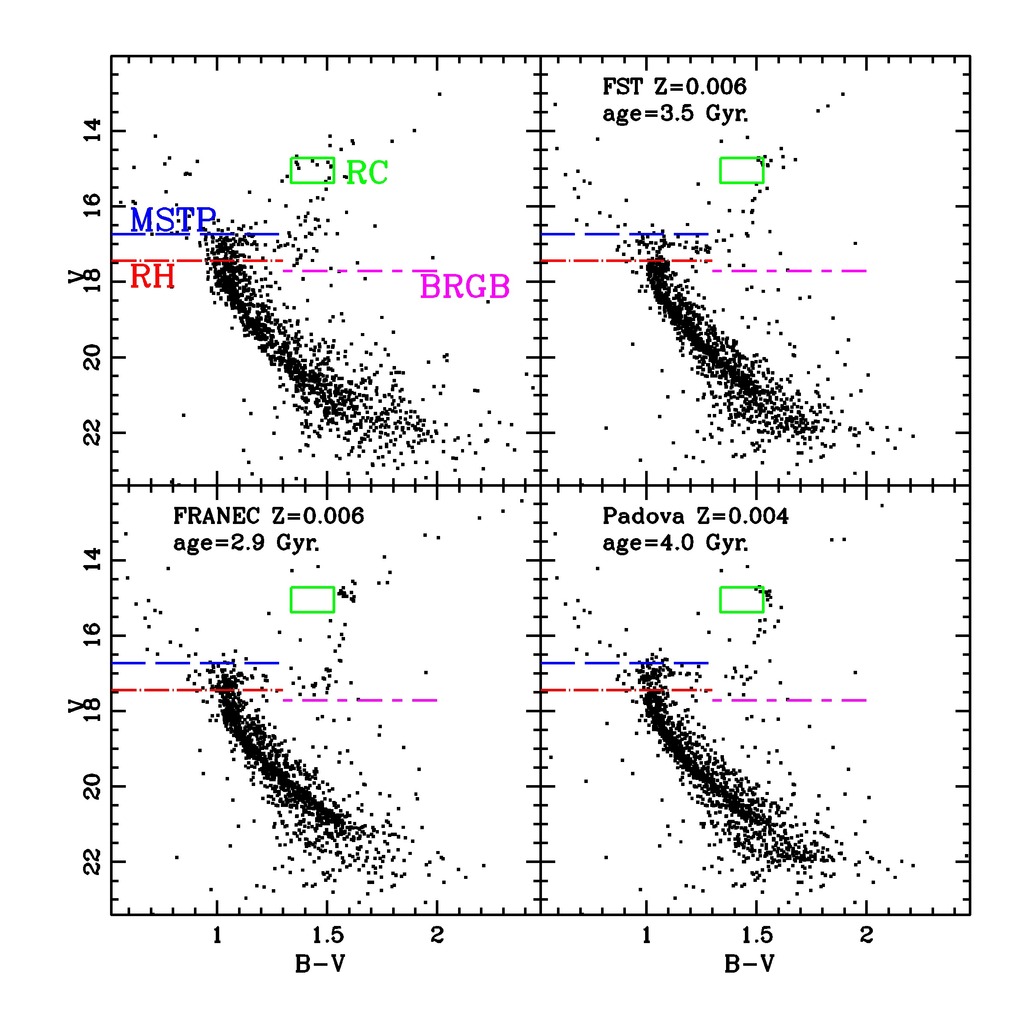}
\caption{Upper left panel: DR-corrected observational CMD. The other panels are the synthetic CMDs for the different evolutionary models used and described in this section.}
\label{figsynths}
\end{figure*}

Figure~\ref{figsynths} shows the comparison between the DR-corrected observational CMD inside 3$\arcmin$ from the cluster centre (top left) and the best fits obtained with the three sets of tracks. To better compare data and models we indicated the age sensitive indicators (see Sec.~\ref{sec:CMD}) both in the observed and in the synthetic CMDs.

For the Padova models, the set in better agreement with the observational features has metallicity $Z=0.004$ ([Fe/H]$\sim-0.6$ dex). We were able to well reproduce the MS shape, the RH and MSTP colours and luminosity levels. The age is $4.0\pm0.2$ Gyr. We obtained a slightly redder RGB and RC in the synthetic CMDs (less than 0.1 mag) but the luminosity levels of both RC and BRGB are reproduced. The estimates for reddening and distance modulus for this solution are $E(B-V)=0.62\pm0.04$ mag and $(m-M)_0=12.25\pm0.1$ mag.

In the case of the FRANEC models the best tracks have $Z=0.006$ ([Fe/H]$\sim-0.5$ dex). We found a general good fit of the luminosity level for the RH, MSTP, BRGB, and RC indicators, but generally a redder colour for
RGB and RC (about 0.1 mag). We obtained the best fit for an age of $2.9\pm0.2$ Gyr (models without overshooting always predict smaller age values), a reddening of $E(B-V)=0.66\pm0.04$ mag, and a distance modulus $(m-M)_0=12.44\pm0.1$ mag. 

The best agreement between synthetic and observed CMDs is found with the FST models with moderate overshooting, in terms of the best reproduction of the overall observed CMD morphology. The differences of the synthetic CMDs for $Z=0.006$ ([Fe/H]$\sim-0.5$ dex) are very small. The MS shape is well reproduced as well as the luminosity level of the age sensitive indicators RH, MSTP, BRGB and RGB. Also in this case the colour of the giant phases are slightly redder but by less than 0.1 mag. The parameter estimates are: age $3.5\pm0.2$ Gyr, $E(B-V)=0.60\pm0.04$ mag, and $(m-M)_0= 12.35\pm0.1$ mag.

Table~\ref{tab:summary} shows the cluster parameters we derived, together with
the implied Galactocentric distance and height.
We recall that literature values range between 3 and 6 Gyr for the age, 0.5 and 0.8 mag for the reddening, 10 and 12.6 mag for the distance modulus and there is general agreement on a subsolar metallicity. The differences are mostly due to the methods used by the authors 
in deriving the parameters. For example, K98 used the morphological parameters $\Delta V$ and/or 
$\Delta (B-V)$, $\Delta (V-I)$ \citep[e.g.,][]{castellani} of the magnitude/colour differences between TO 
and RC. They estimated an age of 4.1 Gyr but adopting solar metallicity. P04 used a similar method, the $\delta V$ index - the 
difference in magnitude between the RC and the TO level \citep{pjm94} - and the Morphological Age Index 
(MAI) calibrated by \cite{jp94} estimating an age of 4.6 Gyr. However these methods are weak when one or both the TO or RC phases on the CMD are contaminated by field stars. This is the case 
of Tr~5: although the RC is evident, the TO is not, even in the inner parts of the cluster. Our method uses 
at the same time more age sensitive indicators together, hence it is more robust against such 
uncertainties.

\begin{table*}
  \centering
  \caption{Cluster parameters derived using different models. Recall that the spectroscopic metallicity we found is [Fe/H]$=-0.4$.}
  \begin{tabular}{|l|c|c|c|c|c|c|c|c|}
    \hline
    \hline
 Model & age & $Z^a$ & $(m-M)_0$ & $E(B-V)$ & $d_{\odot}$ & $R_{GC}^b$ &Z & $M_{TO}$\\
       & (Gyr) &     & (mag)     & (mag)    & (kpc)       & (kpc)      & (pc) & ($M_{\odot}$)\\
    \hline
 Padova & 4.0  & 0.004  & 12.25 & 0.62 & 2.82 & 10.65 & 49.9 & 1.15 \\
 FST    & 3.5  & 0.006  & 12.35 & 0.60 & 2.95 & 10.78 & 52.3 & 1.24 \\
 FRANEC & 2.9  & 0.006  & 12.44 & 0.66 & 3.08 & 10.90 & 54.5 & 1.23 \\ 
    \hline
  \multicolumn{9}{l}{$^a$Metal content of the evolutionary tracks.}\\
  \multicolumn{9}{l}{$^bR_{\odot}=8$ kpc is used to compute $R_{GC}$.}\\
  \end{tabular}
  \label{tab:summary}
\end{table*}

\section{Summary and discussion} \label{summary} 
Tr~5 is a populous OC in the anticentre direction. We were able to perform a complete analysis of the cluster properties by combining the information from photometric and spectroscopic observations. 

We  obtained a CMD about two mag deeper than the literature ones and on a larger FoV thanks to the WFI instrument. 
We found that the cluster is located in a region with large differential reddening and we were able to quantify it within a radius of 
$4\arcmin$. The synthetic CMD technique allowed us to infer the most likely range for age, metallicity, binary fraction, reddening, and distance. We used three different sets of stellar tracks (Padova, FST, FRANEC) to describe the evolutionary status of the cluster in order to take into account how different models impact the accuracy of the analysis. We found that Tr~5 is located at about 3 kpc from the Sun. Its position in the Galactic disc is at $R_{GC}\sim11$ kpc and 50 pc above the plane (assuming $R_\odot=8$ kpc). The resulting age is
between 2.9 and 4 Gyr, depending on the adopted stellar model, with better fits for ages between 3.5 and 4.0 Gyr. The mean Galactic reddening $E(B-V)$ is between 0.6 and 0.7 mag and we estimate a fraction of binaries of at least 25\%. The photometric  metallicity is lower than solar, in the range $0.004<Z<0.006$. While
this estimation is coarse,  the agreement with the spectroscopic analysis is very good.

\begin{figure}
\centering
\includegraphics[width=8.5cm,height=15cm]{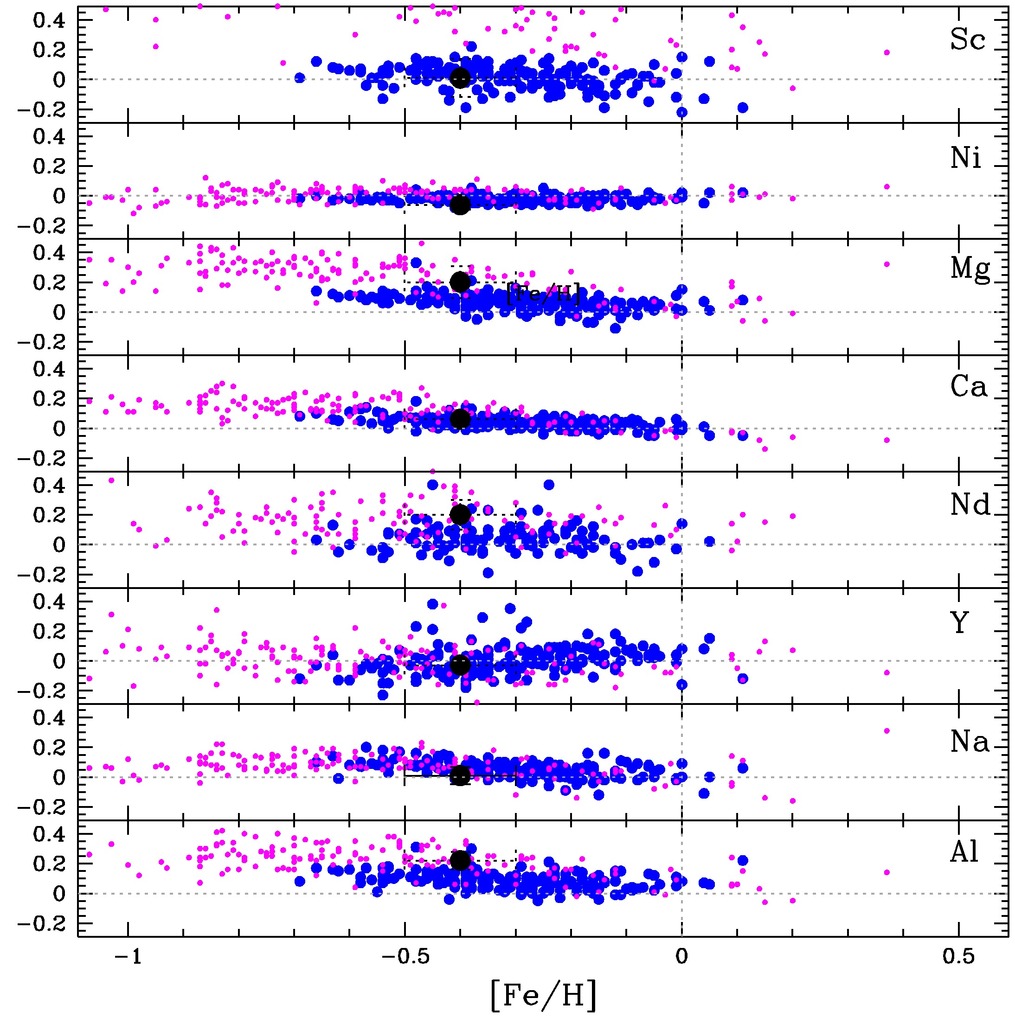}
\caption{Comparison between the abundances derived for two light elements (Na
and Al), two alpha-elements (Mg and Ca), two fe-peak elements (Sc and Ni), and
to heavy elements (Y and Nd) for Tr~5 (black dot) with thin (blue,
\citealt{reddy03}) and thick (magenta, \citealt{reddy06}) disc field stars. }
\label{figfield}
\end{figure}

From the analysis of the high-resolution UVES spectra of three RC stars  we
derived an average cluster metallicity [Fe/H]=$-0.403\pm0.006$~dex, i.e., with a
tiny spread. This was also found by a recent analysis of four other RC stars by
\citet{monaco}. We also derived abundances of light, $\alpha$, Fe-group, and
n-capture elements All their ratios to iron appear to be solar with a very
small  dispersion within the cluster (see Table~\ref{tababu}). 
The abundance ratios we obtained are
typical of thin disc star of similar metallicity, as shown also by
Fig.~\ref{figfield}, where we plot the results for Tr~5 together with those of
field disc stars, both of the thin and the thick component
\citep{reddy03,reddy06}. In general, the obtained abundances for Tr~5 match
those observed in the thin disc. For some chemical species like Mg, Nd, and Al
the observed abundance ratios seem more similar to those observed in the thick
disc; however, taking into account the errorbars, there is no real discrepancy
with the thin disc's distribution.   

\begin{figure}
\centering
\includegraphics[scale=0.9]{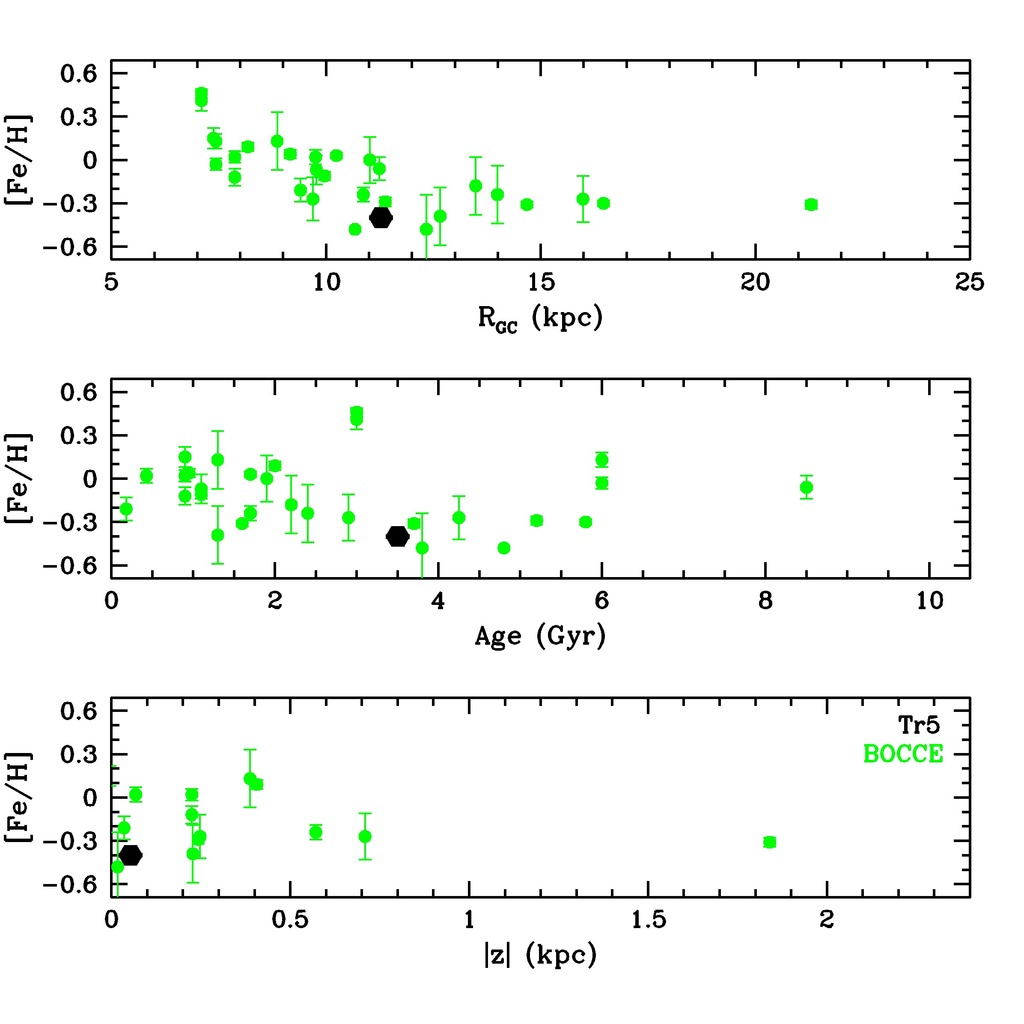}
\caption{Run of [Fe/H] with R$_{GC}$ (top), age (middle), and $|z|$ (bottom) of
the OCs studied by BOCCE (green points). Tr~5 is the black hexagon. Note that
the error bars are smaller than the size of the point.}
\label{figtrend}
\end{figure}

In Figure~\ref{figtrend} we plotted the location of Tr~5 in the R$_{GC}$, age, and distance from the plane versus [Fe/H]. As comparison we also plotted other OCs in the BOCCE sample using metallicity determined from high resolution spectroscopy, either by our group, or in literature papers. Tr~5 does not show any peculiar property, but it is clearly one of the oldest and most metal-poor OCs. It is located near the Galactocentric distance where the radial metallicity gradient changes slope and flattens (see upper panel of Fig.~\ref{figtrend}) where there are other OCs with similar features. This picture is independent of the OCs metallicity source used  (see e.g. \citealt{heiter}). More homogeneous OC metallicities and fundamental parameters will be soon available thanks to the the results expected from the Gaia satellite and the on-going large spectroscopic surveys, like APOGEE \citep{majewski} and the Gaia-ESO Survey \citep{gilmore,randich}, which are targeting many OCs, see e.g. the first results presented by \cite{frinchaboy} for the former and \cite{donati14b,friel,magrini} for the latter. Anyway, detailed studies combining photometry, spectroscopy and evolutionary models for particular interesting clusters, like Tr~5, will still be needed in the large surveys era.

\section*{Acknowledgements}

We thank Paolo Montegriffo (INAF-Osservatorio Astronomico di Bologna, Italy) for his software CataPack, Alessio Mucciarelli for his help with GALA, and Michele Cignoni for useful discussions. This research has made use of NASA's Astrophysics Data System, the SIMBAD database and Aladin (operated at CDS, Strasbourg, France), and the WEBDA database (originally developed by J.-C. Mermilliod, now operated at the Department of Theoretical Physics and Astrophysics of the Masaryk University, Brno). This research has been partially funded by MIUR and INAF (grants ``The Chemical
and Dynamical Evolution of the Milky Way and Local Group Galaxies'' ,
prot. 2010LY5N2T; grant ``Premiale VLT 2012").
DSS was produced at the Space Telescope Science Institute under U.S. Government
grant NAGW-2166. The images of these surveys are based
on photographic data obtained using the Oschin Schmidt Telescope
on Palomar Mountain and the UK Schmidt Telescope. The plates
were processed into the present compressed digital form with the
permission of these institutions), and the Guide Star Catalogue-II
(GSC II is a joint project of the Space Telescope Science Institute)
and the Osservatorio Astronomico di Torino. Funding for the SDSS and SDSS-II has been provided by the Alfred P. Sloan Foundation, the Participating Institutions, the National Science Foundation, the U.S. Department of Energy, the National Aeronautics and Space Administration, the Japanese Monbukagakusho, the Max Planck Society, and the Higher Education Funding Council for England. The SDSS Web Site is http://www.sdss.org/.
The SDSS is managed by the Astrophysical Research Consortium for the Participating Institutions. The Participating Institutions are the American Museum of Natural History, Astrophysical Institute Potsdam, University of Basel, University of Cambridge, Case Western Reserve University, University of Chicago, Drexel University, Fermilab, the Institute for Advanced Study, the Japan Participation Group, Johns Hopkins University, the Joint Institute for Nuclear Astrophysics, the Kavli Institute for Particle Astrophysics and Cosmology, the Korean Scientist Group, the Chinese Academy of Sciences (LAMOST), Los Alamos National Laboratory, the Max-Planck-Institute for Astronomy (MPIA), the Max-Planck-Institute for Astrophysics (MPA), New Mexico State University, Ohio State University, University of Pittsburgh, University of Portsmouth, Princeton University, the United States Naval Observatory, and the University of Washington.

\label{lastpage}


\begin{thebibliography}{}

\bibitem[\protect\citeauthoryear{Alonso, Arribas, \& Mart{\'{\i}}nez-Roger}{1999}]{alonso99} 
Alonso A., Arribas S., Mart{\'{\i}}nez-Roger C., 1999, A\&AS, 140, 261 

\bibitem[\protect\citeauthoryear{Ahumada et al.}{2013}]{ahumada13} 
Ahumada A.~V., Cignoni M., Bragaglia A., 
Donati P., Tosi M., Marconi G., 2013, MNRAS, 430, 221 

\bibitem[\protect\citeauthoryear{Bessell, Castelli, 
\& Plez}{1998}]{bessel98} Bessell M.~S., Castelli F., Plez B., 1998, A\&A, 333, 231 

\bibitem[\protect\citeauthoryear{Bonnarel et al.}{2000}]{aladin} 
Bonnarel F., et al., 2000, A\&AS, 143, 33 

\bibitem[\protect\citeauthoryear{Bragaglia \& Tosi}{2006}]{bt06}
Bragaglia A., Tosi M., 2006, AJ, 131, 1544 

\bibitem[\protect\citeauthoryear{Bragaglia et  al.}{2008}]{bragaglia08}
Bragaglia A., Sestito P., Villanova S., Carretta E., Randich S., Tosi M., 2008,
A\&A, 480, 79 

\bibitem[\protect\citeauthoryear{Bressan et al.}{1993}]{bbc93} 
Bressan A., Fagotto F., Bertelli G., Chiosi C., 1993, A\&AS, 100, 647 

\bibitem[\protect\citeauthoryear{Brogaard et al.}{2012}]{brogaard} 
Brogaard K., et al., 2012, A\&A, 543, A106 

\bibitem[\protect\citeauthoryear{Cantat-Gaudin et al.}{2014}]{doop} 
Cantat-Gaudin T., et al., 2014, A\&A, 562, A10 

\bibitem[\protect\citeauthoryear{Carrera et al.}{2007}]{carrera07} 
Carrera R., Gallart C., Pancino E., Zinn  R., 2007, AJ, 134, 1298 

\bibitem[\protect\citeauthoryear{Carrera \& Pancino}{2011}]{carrera11} 
Carrera R., Pancino E., 2011, A\&A, 535, A30 

\bibitem[\protect\citeauthoryear{Castellani et al.}{2000}]{castellani} 
Castellani V., Degl'Innocenti S., Girardi L., Marconi M., Prada Moroni P.~G., Weiss A., 2000, A\&A, 354, 150 

\bibitem[\protect\citeauthoryear{Castelli \& Kurucz}{2003}]{castelli} 
Castelli F., Kurucz R.~L., 2003, IAUS, 210, 20P 

\bibitem[\protect\citeauthoryear{Cayrel et al.}{2004}]{cayrel} 
Cayrel R., et al., 2004, A\&A, 416, 1117 

\bibitem[\protect\citeauthoryear{Cignoni et al.}{2011}]{cignoni11} 
Cignoni M., Beccari G., Bragaglia A., Tosi M., 2011, MNRAS, 416, 1077

\bibitem[\protect\citeauthoryear{Cole et al.}{2004}]{cole04} 
Cole A.~A., Smecker-Hane T.~A., Tolstoy E., Bosler T.~L., Gallagher J.~S., 
2004, MNRAS, 347, 367 

\bibitem[\protect\citeauthoryear{D'Odorico et al.}{2000}]{dekker} 
D'Odorico S., Cristiani S., Dekker H., Hill V., Kaufer A., Kim T., Primas F., 2000, SPIE, 4005, 121 

\bibitem[\protect\citeauthoryear{Dean, Warren, \& Cousins}{1978}]{dean_78} Dean J.~F., Warren P.~R., Cousins A.~W.~J., 1978, MNRAS, 183, 569 

\bibitem[\protect\citeauthoryear{Dias et al.}{2002}]{dias02} 
Dias W.~S., Alessi B.~S., Moitinho A., L{\'e}pine J.~R.~D., 2002, A\&A, 389, 871 

\bibitem[\protect\citeauthoryear{Dominguez et al.}{1999}]{franec} 
Dominguez I., Chieffi A., Limongi M., Straniero O., 1999, ApJ, 524, 226 

\bibitem[\protect\citeauthoryear{Donati et al.}{2012}]{donati12} 
Donati P., Bragaglia A., Cignoni M., Cocozza G., Tosi M., 2012, MNRAS, 424, 
1132 

\bibitem[\protect\citeauthoryear{Donati et al.}{2014a}]{donati14a} 
Donati P., Beccari G., Bragaglia A., Cignoni M., Tosi M., 2014a, MNRAS, 437, 
1241 

\bibitem[\protect\citeauthoryear{Donati et al.}{2014b}]{donati14b} 
Donati P., et al., 2014b, A\&A, 561, A94 

\bibitem[\protect\citeauthoryear{Dow \& Hawarden}{1970}]{dh70} 
Dow M.~J., Hawarden T.~G., 1970, MNSSA, 29, 137 

\bibitem[\protect\citeauthoryear{Fagotto et al.}{1994}]{fag94} Fagotto F., Bressan A., Bertelli G., Chiosi C., 1994, A\&AS, 105, 29 

\bibitem[\protect\citeauthoryear{Friel}{1995}]{friel95} 
Friel E.~D., 1995, ARA\&A, 33, 381 

\bibitem[\protect\citeauthoryear{Friel et 
al.}{2014}]{friel} Friel E.~D., et al., 2014, A\&A, 563, A117 

\bibitem[\protect\citeauthoryear{Frinchaboy et 
al.}{2013}]{frinchaboy} Frinchaboy P.~M., et al., 2013, ApJ, 777, 

\bibitem[\protect\citeauthoryear{Gilmore et al.}{2012}]{gilmore} 
Gilmore G., et al., 2012, Msngr, 147, 25 

\bibitem[\protect\citeauthoryear{Grevesse, Noels, \& Sauval}{1996}]{grevesse96}
Grevesse N., Noels A., Sauval A.~J., 1996, ASPC, 99, 117 

\bibitem[\protect\citeauthoryear{Gustafsson et  al.}{2008}]{marcs} 
Gustafsson B., Edvardsson B., Eriksson K., J{\o}rgensen U.~G., Nordlund {\AA}.,
Plez B., 2008, A\&A, 486, 951 

\bibitem[\protect\citeauthoryear{Heiter et al.}{2014}]{heiter} 
Heiter U., Soubiran C., Netopil M., Paunzen E., 2014, A\&A, 561, A93 

\bibitem[\protect\citeauthoryear{Janes \& Phelps}{1994}]{jp94} 
Janes K.~A., Phelps R.~L., 1994, AJ, 108, 1773 

\bibitem[\protect\citeauthoryear{Kaluzny}{1998}]{kaluzny98} 
Kaluzny J., 1998, A\&AS, 133, 25 

\bibitem[\protect\citeauthoryear{Kim \& Sung}{2003}]{ks03} 
Kim S.~C., Sung H., 2003, JKAS, 36, 13 

\bibitem[\protect\citeauthoryear{Kim, Kyeong, \& Sung}{Kim et al.}{2009}]{kks09} 
Kim S.~C., Kyeong J., Sung E.-C., 2009, JKAS, 42, 135 

\bibitem[\protect\citeauthoryear{Koch et al.}{2003}]{koch03} 
Koch A., Odenkirchen M., Caldwell J.~A.~R., Grebel E.~K., 2003, ANS, 324, 
95 

\bibitem[\protect\citeauthoryear{Kurucz}{2005}]{kurucz} 
Kurucz, R.~L.\ 2005, Memorie  della Societa Astronomica Italiana Supplementi, 8, 14 

\bibitem[\protect\citeauthoryear{Landolt}{1992}]{landolt}
Landolt A., 1992, AJ, 104, 340

\bibitem[\protect\citeauthoryear{L{\'e}pine et al.}{2011}]{lepine11} 
L{\'e}pine J.~R.~D., et al., 2011, MNRAS, 417, 698 

\bibitem[\protect\citeauthoryear{Magrini et al.}{2009}]{magrini09} 
Magrini L., Sestito P., Randich S., Galli D., 2009, A\&A, 494, 95 

\bibitem[\protect\citeauthoryear{Magrini et 
al.}{2014}]{magrini} Magrini L., et al., 2014, A\&A, 563, A44 

\bibitem[\protect\citeauthoryear{Majewski et 
al.}{2010}]{majewski} Majewski S.~R., Wilson J.~C., Hearty F., 
Schiavon R.~R., Skrutskie M.~F., 2010, IAUS, 265, 480 

\bibitem[\protect\citeauthoryear{Milone et al.}{2012}]{mil12} 
Milone A.~P., et al., 2012, A\&A, 540, A16 

\bibitem[\protect\citeauthoryear{Monaco et al.}{2014}]{monaco} 
Monaco L., et al., 2014, A\&A, 564, L6 

\bibitem[\protect\citeauthoryear{Mucciarelli et al.}{2013}]{gala} 
Mucciarelli A., Pancino E., Lovisi L.,  Ferraro F.~R., Lapenna E., 2013, ApJ, 766, 78 

\bibitem[\protect\citeauthoryear{Pancino et al.}{2010}]{pancino10} 
Pancino E., Carrera R., Rossetti E., Gallart C., 2010, A\&A, 511, A56 

\bibitem[\protect\citeauthoryear{Phelps, Janes, \& Montgomery}{Phelps et al.}{1994}]{pjm94} 
Phelps R.~L., Janes K.~A., Montgomery K.~A., 1994, AJ, 107, 1079 

\bibitem[\protect\citeauthoryear{Piatti, Clari{\'a}, \& Ahumada}{Piatti et al.}{2004}]{piatti04} 
Piatti A.~E., Clari{\'a} J.~J., Ahumada A.~V., 2004, MNRAS, 349, 641 

\bibitem[\protect\citeauthoryear{Platais et al.}{2012}]{platais} 
Platais I., et al., 2012, ApJ, 751, L8 

\bibitem[\protect\citeauthoryear{Randich et al.}{2013}]{randich} 
Randich S., Gilmore G., Gaia-ESO Consortium, 2013, Msngr, 154, 47 

\bibitem[\protect\citeauthoryear{Reddy et al.}{2003}]{reddy03} 
Reddy B.~E., Tomkin J., Lambert D.~L., Allende Prieto C., 2003, MNRAS, 340, 
304 

\bibitem[\protect\citeauthoryear{Reddy, Lambert, 
\& Allende Prieto}{2006}]{reddy06} Reddy B.~E., Lambert D.~L., Allende Prieto C., 2006, MNRAS, 367, 1329 

\bibitem[\protect\citeauthoryear{Ryabchikova, Pakhomov, \& Piskunov}{2011}]{vald3}
Ryabchikova T.~A., Pakhomov Y.~V., Piskunov N.~E., 2011, KIzKU, 153, 61 

\bibitem[\protect\citeauthoryear{Sbordone et al.}{2004}]{sbordone} 
Sbordone, L.,  Bonifacio, P., Castelli, F.,  \& Kurucz, R.~L.\ 2004, Memorie
della Societa Astronomica Italiana Supplementi, 5, 93 

\bibitem[\protect\citeauthoryear{Sestito et  al.}{2008}]{sestito08} 
Sestito P., Bragaglia A., Randich S., Pallavicini R., Andrievsky S.~M., Korotin
S.~A., 2008, A\&A, 488, 943 

\bibitem[\protect\citeauthoryear{Skrutskie et al.}{2006}]{2mass} 
Skrutskie M.~F., et al., 2006, AJ, 131, 1163 

\bibitem[\protect\citeauthoryear{Stetson}{1987}]{ste1}
Stetson P. B., 1987, PASP, 99, 191

\bibitem[\protect\citeauthoryear{Stetson}{1993}]{ste2}
Stetson P. B., 1993, in Butler C. J., Elliott I., eds, Stellar photometry -
Current techniques and future developments, Proc. IAU Colloquium
136, Cambridge Univ. Press, Cambridge, p. 291

\bibitem[\protect\citeauthoryear{Stetson \& Pancino}{2008}]{daospec} 
Stetson P.~B., Pancino E., 2008, PASP, 120, 1332 

\bibitem[\protect\citeauthoryear{Tosi et al.}{1991}]{tosi91} 
Tosi M., Greggio L., Marconi G., Focardi P., 1991, AJ, 102, 951

\bibitem[\protect\citeauthoryear{Ventura et al.}{1998}]{fst} 
Ventura P., Zeppieri A., Mazzitelli I., D'Antona F., 1998, A\&A, 334, 953 

\bibitem[\protect\citeauthoryear{Yong, Carney, \& Friel}{Yong et
al.}{2012}]{yong12}  Yong D., Carney B.~W., Friel E.~D., 2012, AJ, 144, 95 


\end{thebibliography}
\end{document}